\documentclass[12pt,a4paper]{article}
\usepackage[margin=1in]{geometry}
\usepackage[T1]{fontenc}
\usepackage[utf8]{inputenc}
\usepackage{lmodern}
\usepackage{amsmath,amssymb,amsfonts,amsthm,mathtools,bm,mathrsfs}
\usepackage{graphicx,booktabs,tabularx,longtable,multirow,makecell}
\usepackage{algorithm,algpseudocode}
\usepackage{natbib}
\usepackage{xcolor,pifont,microtype,xurl}
\usepackage[normalem]{ulem}
\usepackage{hyperref}
\hypersetup{colorlinks=true,linkcolor=blue,citecolor=blue,urlcolor=blue}
\newtheorem{example}{Example}

\newcommand{\R}{\mathbb{R}}

\title{Local Variable Selection for Spatial Point Process with Application in Violent Crime Modeling}
\author{Debjoy Thakur$^{1,2,*}$ \qquad Soumendra N. Lahiri$^{1}$\\[0.7em]
{\small $^{1}$Department of Statistics and Data Science, Washington University in St. Louis, Missouri, USA}\\
{\small $^{2}$Ahmedabad University, India}\\[0.35em]
{\small $^{*}$Corresponding author: \href{mailto:debjoythakur@outlook.com}{debjoythakur@outlook.com, debjoy.thakur@ahduni.edu.in}}\\
{\small ORCID (Debjoy Thakur): \href{https://orcid.org/0000-0001-5514-5707}{0000-0001-5514-5707}}}
\date{}
\begin{document}
\maketitle
\begin{abstract}
In recent years, researchers have increasingly considered penalized methods for estimating inhomogeneous intensity functions in spatial point processes in the presence of a large number of predictors using different types of convex and non-convex penalty function. However, these variable selection methods select the variables globally instead of capturing spatial heterogeneity in the set of selected variables. To fill the gap, we have proposed a novel local-level variable selection method for spatial point process data. Its advantage is twofold: it not only selects the predictors but also provides information about location-specific variable selection. Our research is motivated by crime occurrences in  St. Louis city which identifies the incidence point specific set of relevant predictors. We describe the simulation results to justify the accuracy of local-level variable selection during intensity function estimation.
\end{abstract}
\par
\medskip
\noindent\textbf{Key messages}
\begin{itemize}
\item We identify the location-specific selected variables.
\item We have identified that there is a high concentration of violent crimes in the central and southern part of St. Louis and high-concentration of bars, buildings, and public transportation variables in the neighborhood are the driving factors.
\item Local LASSO based regularization model becomes the most parsimonious model in this real data.
\end{itemize}
\section{Introduction}
Urban safety of a city mainly relies on its crime occurrences. St. Louis, Missouri is recognized as one of the deadliest cities with the highest crime rates, especially in the context of violent crimes. The city generally involves significant levels of violent crimes, for example, homicide, robbery, and aggravated assaults, and so on. In general, crime rates are relatively higher than in other American cities of the same size. The high concentration of violent crimes risks public safety and quality of life in the city. The city of St. Louis stands in the third position in the year 2025 according to the recent report, which is accessible at the following link: \url{https://www.noradarealestate.com/blog/top-10-most-dangerous-cities-in-america/\#FBIS_TOP_10_MOST_DANGEROUS_CITIES_IN_THE_MIDWEST}. Traditionally, we observe that these crimes are concentrated where a huge number of poorly lit areas and fragmented buildings are clustered. These occurrences of crime across space lead to considerable difficulties and possibilities for statistical modeling. According to \cite{mohler2019reducing}, spatial and temporal crime clusters emerge in urban environments, and the occurrence of one event induces subsequent crime or accident events through retaliation or repeated victimization (\cite{heckman1991identifying}). A central difficulty for empirical analysis is that the co-occurrences of retaliation and dependence hardly identify the reason for clusters in observational data (\cite{diggle2013statistical}).

An increasing number of studies have emphasized the use of spatial and spatio-temporal point process models to analyze and forecast the dynamics of urban crime and its corresponding hazards. Early work by \cite{mohler2011self} has described how criminal activity increases the likelihood of future incidents at nearby locations within a short time span, using residential burglary data from Los Angeles. \cite{andresen2012spatial} has scrutinized the sensitivity of crime patterns to the different scales of spatial aggregation. This study reveals that a smaller geographical area substantially captures the local variability better than the coarser-level analysis, which increases the importance of spatial resolution during modeling localized crime risk. Subsequently, \cite{mohler2014marked} has extended the point process formulation to a marked framework that jointly models multiple crime types and captures both short-range and long-range dependent patterns of risk. \cite{reinhart2018self} has developed a spatio-temporal self-exciting model that integrates spatial covariates under the assumption that events are repeated and retaliatory behavior occurs. This study has highlighted how spatial inhomogeneity and temporal triggering are often intertwined. More recently, \cite{chen2025score}'s Neural Temporal Point Process (NTPP) with a conditional spatial module learns latent dependencies between past and future events to capture complex dependencies in traffic accident data. \cite{park2021investigating} modeled gang-related violent crime in Los Angeles (2014–2017) using a marked spatio-temporal Hawkes process with demographic covariates. They estimate non-parametrically the spatially varying background rate via kernel smoothing and generalized additive models to distinguish structural inhomogeneity from event-driven retaliation. Recently, \cite{van2025non} has proposed a marked point process using a non-homogeneous Markov point process as the ground process for the occurrence of Aoristic events.

Beyond these problems, one persistent problem is selecting potential locally associated features to learn the spatial heterogeneity behind the crime occurrences. From a methodological point of view, the diverging number of predictors forces regularized intensity function estimation for the spatial point process. Traditionally, point‐process modeling has assumed relatively modest numbers of covariates, and has relied on estimating functions based on the Campbell theorem, logistic, or Poisson approximations (\cite{waagepetersen2007estimating}, \cite{waagepetersen2008estimating}, \cite{waagepetersen2009two}, \cite{baddeley2014logistic}). More recently, however, the explosion of available spatial data enables high‐dimensional settings where the number of features may exceed the number of observed events (\cite{renner2013equivalence}, \cite{thurman2014variable}, \cite{thurman2015regularized}, \cite{choiruddin2023adaptive}, \cite{choiruddin2018convex}). In such cases, variable selection becomes essential. Adaptive LASSO and elastic‐net methods have been applied to inhomogeneous Poisson point‐process intensity estimation (\cite{thurman2015regularized}, \cite{yue2015variable}). \cite{thurman2014variable}, \cite{choiruddin2018convex}, \cite{choiruddin2023adaptive} have discussed the theoretical properties of variable selection, for example, the existence of a local maximizer, sparsity selection, and asymptotic normality, in detail. In the machine learning community, \cite{biscio2025nonparametric} has used random forest for intensity function modeling in the presence of a large number of predictors, \cite{flaxman2017poisson} has proposed a new Reproducing Kernel Hilbert Space formulation for the inhomogeneous intensity function, where they model the square root of the intensity function of the inhomogeneous Poisson process using the penalized likelihood and later \cite{kim2022fast} has extended \cite{flaxman2017poisson}'s approach by proposing an augmented permanetal process (APP). Recently, \cite{lu2026xgboostpp} has numerically demonstrated a deterioration in APP performance with increasing numbers of predictors and introduced a new tree-based approach for the non-parametric estimation of the inhomogeneous intensity function in the high-dimensional setting. In this article, we extend this literature by combining two strands: 
\begin{enumerate}
    \item First, we employ Haar-wavelet basis expansion for the spatially varying coefficient function to incorporate the localization in the likelihood of inhomogeneous spatial point process.
    \item Secondly, we select the relevant predictors at the local level. This allows us to detect not only whether a predictor matters but also where it matters.
\end{enumerate}
  In Section~\ref{s:mot} we describe the motivational social problem behind this research and clearly distinguish our contributions with respect to the previous research. Section~\ref{s:method} describes the methodological and computational details of regularized intensity function estimation with the help of two-dimensional Haar wavelet basis functions. Finally, in Section~\ref{s:sim} we summarize the results regarding the simulation study, and in Section~\ref{s:real} we discuss the local variable selection for St. Louis city and also talk about estimating the intensity function of crime occurrences based on selected predictors at the local level and compare the performance of our methods with Adaptive Lasso (AL), Lasso, and SCAD regularized intensity function estimation methods. 
\section{Motivation and Contribution}\label{s:mot}
\begin{figure}[h!]
    \centering
    \includegraphics[width=\linewidth]{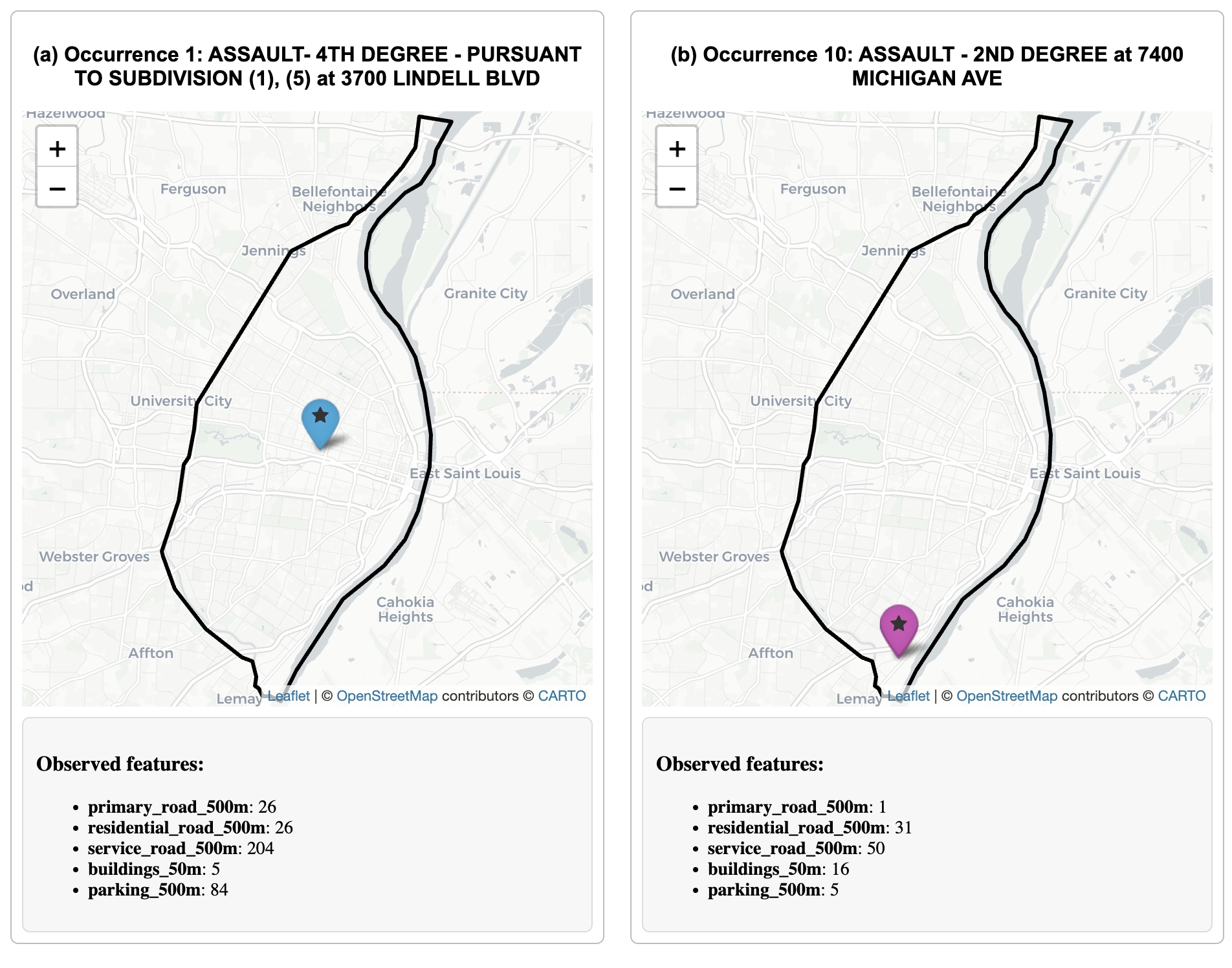}
    \caption{Motivation behind Local Variable Selection}
    \label{fig:motivation}
\end{figure}

In this section, we describe the motivation behind local-variable selection during intensity surface estimation. we will emphasize the importance of localized risk pattern modeling for variable selection by presenting a real example in Figure~\ref{fig:motivation}.
As seen in Figure~\ref{fig:motivation}, we have represented different types of economic features around the occurrences of crime events. For example, we have $5$ buildings within $50$m radius, $26$ residential roads within $500$m radius, $204$ servicing roads within $500$m radius, and $84$ parking points within $500$m radius around $3700$ Lindell Blvd in the central part of St. Louis. Whereas we have $16$ buildings within a $ 50$ m radius, $31$ residential roads within a $ 500$ m radius, $50$ servicing roads within a $ 500$ m radius, and $5$ parking points within a $ 500$ m radius around of $7400$ Michigan Ave in the Southern part of St. Louis. This example establishes the existence of spatial heterogeneity in the location-specific features, such as, number of buildings within $50$m, $200$m, and $500$m radius; number of different types of roads, number of parking areas within different levels of radius; and so on. This discussion expedites the need for localized modeling of intensity function estimation for the occurrence of violent crimes. Hence, the main scientific question is 
\begin{itemize}
    \item ``How can we select location-specific relevant variables ?'' 
\end{itemize}
Therefore, we focus on local-level analysis of variable selection by localizing the coefficient surface. Previously in the literature \cite{baddeley2017local}, \cite{coeurjolly2022convolution}, \cite{yin2021fused} have proposed a spatially varying type coefficient function in the inhomogeneous spatial point process. Although their goal and approach strongly differ from our targets. \cite{baddeley2017local} has already proposed a local composite likelihood for a spatial point process. But this localization happens for a fixed reference point. As a result, whenever we are going to localize around all observed points we need to replicate their modeling approach over and over. \cite{coeurjolly2022convolution} has assumed the coefficients and covariates as a periodic function where Fourier coefficients play a crucial role. But in our framework, we neither assume the periodic property of the coefficient function nor that the Fourier coefficients have the localization feature. They have also used the adaptive lasso type penalty motivated by \cite{choiruddin2018convex}, though their goal is to incorporate sparsity depending on the relevant and irrelevant Fourier coefficients. From this idea, we definitely select relevant Fourier coefficients, but can not justify the selection of a predictor at the local-level. Similarly, \cite{yin2021fused} assumed that the coefficient function is spatially clustered, such that the coefficient function is constant in the same spatial cluster and it varies across different clusters. They have used a graph-based fused penalty term in the penalized negative composite likelihood, which guarantees the homogeneity of the coefficient function over the same cluster. Their target is maintaining the homogeneity of the coefficient function within the same cluster instead of localized variable selection. In this research, our modeling approach seems to be a spatially varying inhomogeneous intensity function at first glance, but the goal is selecting variables locally, for which the Haar wavelet is required; otherwise, we require to replicate \cite{baddeley2017local}'s method many times.
\section{Methods}\label{s:method}
 In this section, we propose a local-level variable selection method for a spatial point process which will give us an idea about the location-specific relevant predictors.
\subsection{Multi-resolution Wavelet Decomposition}
\label{subsec:mra-2d}
A wavelet system is built from two generators: the \emph{father wavelet}, which is known as the scaling function, $\phi$, which captures coarse structure; and the \emph{mother wavelet} $\psi$, which captures local fluctuations. For a detailed description, please go through \cite{mallat1999wavelet}. Suppose $V_j\subset \mathcal L_2(0,1)$ denote the subspace at resolution $j$, spanned by a scaling function, $\phi$, and let $W_j$ be the corresponding detail space spanned by wavelet basis function, $\psi$. Let's define the relation between the approximation and detail subspaces as $V_{j+1} \;=\; V_j \oplus W_j$ ($\oplus$ denotes direct sum) along with $V_j \perp W_j$ i.e. $W_j$ is orthogonal complement to $V_j$ in $V_{j+1}$. This relation induces a nested sequence, $\cdots \subset V_{j-1}\subset V_j\subset V_{j+1}\subset\cdots$ such that $\overline{\bigcup_{j} V_j} = \mathcal L_2(0,1)$. In two dimensions, the approximated subspace, $V_j^{(2)}:=V_j\otimes V_j \subset \mathcal L_2([0,1]^2)$ combines the one-dimensional refinement across both axes:
\[
V_{j+1}^{(2)}
=\underbrace{V_j\otimes V_j}_{\text{scaling}}
\ \oplus\ \underbrace{W_j\otimes V_j}_{\mathcal{W}_j^{(1)}}
\ \oplus\ \underbrace{V_j\otimes W_j}_{\mathcal{W}_j^{(2)}}
\ \oplus\ \underbrace{W_j\otimes W_j}_{\mathcal{W}_j^{(3)}},
\]
where $\otimes$ denotes tensor product. 

In general Haar wavelet produce one scaling family and three detailed wavelet families in three directions: horizontal, vertical, and diagonal. At resolution $j$, each basis function is defined on a block of size $2^{-j}\!\times 2^{-j}$, so larger $j$ means finer localization, and smaller $j$ implies coarser resolution. Here $\Phi_{j,(k_1,k_2)}$ represents the scaling wavelet function at resolution ``$j$'' where $k_1, k_2 \in \left\{0, \dots, 2^{j}-1\right\}$. The other wavelet atoms: $\Psi^{(2)}_{j,(k_1,k_2)}$ indicates the wavelet family in the horizontal direction, $\Psi^{(3)}_{j,(k_1,k_2)}$ indicate the wavelet family in the vertical direction, and $\Psi^{(4)}_{j,(k_1,k_2)}$ indicate the wavelet family in the diagonal direction. Let $\phi_{j,k}(u)=2^{j/2}\phi(2^j u-k)$ and $\psi_{j,k}(u)=2^{j/2}\psi(2^j u-k)$ then the basis atoms on spatial domain, $\mathcal{R}$, assuming $(u_1,u_2)$ is the position of a spatial point, are defined as follows:
\begin{equation}\label{eq:tensor_prod}
\scriptsize
    \begin{split}
   \Phi_{j,(k_1,k_2)}(u_1,u_2)=2^{\,j}\,\Phi(2^j u_1-k_1,\,2^j u_2-k_2) = 2^j \phi(2^j u_1 - k_1) \phi(2^ju_2-k_2) = \phi_{j,k_1}(u_1) \otimes \phi_{j,k_2}(u_2),\\
    \Psi^{(2)}_{j,(k_1,k_2)}(u_1,u_2)=2^{\,j}\,\Psi^{(2)}(2^j u_1-k_1,\,2^j u_2-k_2) = 2^j \psi(2^j u_1 - k_1) \phi(2^ju_2-k_2) = \psi_{j,k_1}(u_1) \otimes \phi_{j,k_2}(u_2),\\
   \Psi^{(3)}_{j,(k_1,k_2)}(u_1,u_2)=2^{\,j}\,\Psi^{(3)}(2^j u_1-k_1,\,2^j u_2-k_2) = 2^j \phi(2^j u_1 - k_1) \psi(2^ju_2-k_2) = \phi_{j,k_1}(u_1) \otimes \psi_{j,k_2}(u_2),\\
   \Psi^{(4)}_{j,(k_1,k_2)}(u_1,u_2)=2^{\,j}\,\Psi^{(4)}(2^j u_1-k_1,\,2^j u_2-k_2) = 2^j \psi(2^j u_1 - k_1) \psi(2^ju_2-k_2) = \psi_{j,k_1}(u_1) \otimes \psi_{j,k_2}(u_2),      
    \end{split}
\end{equation}
and $\tilde{\Psi}$ is the concatenation of the basis vectors $(\Phi, \Psi)^\top$.
For an $M\times M$ image with $M=2^J$ at a chosen coarse level $j_0$, the finite-dimensional expansion of the approximation subspace with its basis version is described as:
\[
\begin{aligned}
  V_{J}^{(2)} &= V_{j_0}^{(2)} \oplus \bigoplus_{j=j_0}^{J-1} \left(\mathcal{W}_j^{(1)} \oplus \mathcal{W}_j^{(2)} \oplus \mathcal{W}_j^{(3)}\right)\\
  &=\{\Phi_{j_0,(k_1,k_2)}\}_{k_1,k_2=0}^{2^{j_0}-1}
\ \cup\
\bigcup_{j=j_0}^{J-1}\ \{\Psi^{(2)}_{j,(k_1,k_2)}\}_{k_1,k_2=0}^{2^j-1} \cup \bigcup_{j=j_0}^{J-1}\ \{\Psi^{(3)}_{j,(k_1,k_2)}\}_{k_1,k_2=0}^{2^j-1} \cup \bigcup_{j=j_0}^{J-1}\ \{\Psi^{(4)}_{j,(k_1,k_2)}\}_{k_1,k_2=0}^{2^j-1}.  
\end{aligned}
\]
In the above expansion there are $4^{j_0}$ scaling atoms at the coarse level $j_0$ and $3\cdot 4^{j}$ detail atoms at each level $j$ along the horizontal, vertical, and diagonal directions. A smaller $j_0$ yields fewer scaling coefficients but more detail levels, and a larger $j_0$ the reverse. Any measurable function, $f(\bm s)\in V_J^{(2)}$ can be written as
\begin{equation}\label{eq:function-decompose}
f(\bm s)
= \sum_{\mathbf{k}} \alpha_{j_0,\mathbf{k}}\,\Phi_{j_0,\mathbf{k}}(\bm s)
\;+\;
\sum_{j=j_0}^{J-1}\ \sum_{l=2}^4\ \sum_{\mathbf{k}}
\tilde{\alpha}_{j,l,\mathbf{k}}\,\Psi^{(l)}_{j,\mathbf{k}}(\bm s),
\end{equation}
where coefficients are localized (by $\mathbf{k}=(k_1, k_2)^\top$), and $k_1,k_2$ respectively denote the shift vectors along horizontal and vertical directions. The scale parameter is represented by $j$.
\begin{figure}[!htbp]
    \centering
    \includegraphics[width=\linewidth]{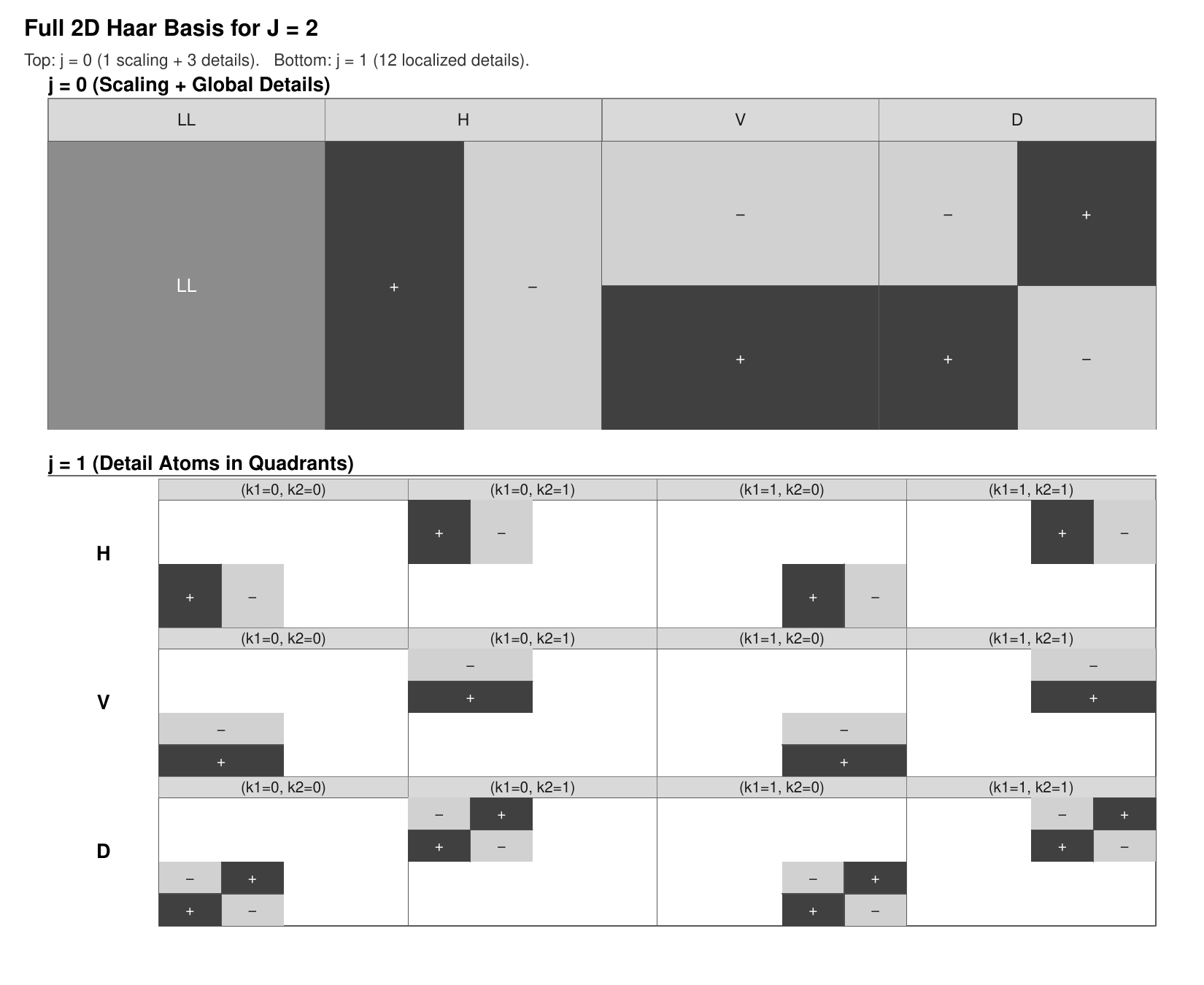}
    \caption{Full 2D Haar basis for $J = 2$. Top row: global scaling and details
$(j = 0)$. Bottom: localized detail atoms at $j = 1$ for H/V/D across four quadrants
$(k_1, k_2) \in \{0, 1\}^2$.}
    \label{fig:localwavelet}
\end{figure}
Figure~\ref{fig:localwavelet} displays the complete two-dimensional Haar basis when $J=2$. Corresponding to $J=2$, the approximated subspace becomes
\[
V_{2}^{(2)}
=V_0^{(2)}\ \oplus\ \mathcal{W}_0^{(1)} \ \oplus\ \mathcal{W}_0^{(2)} \ \oplus\ \mathcal{W}_0^{(3)} \ \oplus\ \mathcal{W}_1^{(1)} \ \oplus\ \mathcal{W}_1^{(2)} \ \oplus\ \mathcal{W}_1^{(3)}.
\]
The \emph{top row} of Figure~\ref{fig:localwavelet} consists of one global scaling atom which spans $V_0^{(2)}$ and three global details (H/V/D) over the entire unit square, which spans $\mathcal{W}_0^{(1)}\oplus \mathcal{W}_0^{(2)}\oplus \mathcal{W}_0^{(3)}$ at $j_0 = 0$. The \emph{bottom grid} shows twelve localized details at $j=1$. For each of the four quadrants $(k_1,k_2)\in\{0,1\}^2$ there are H, V, and D atoms, which span $\mathcal{W}_1^{(1)}\oplus \mathcal{W}_1^{(2)}\oplus \mathcal{W}_1^{(3)}$ corresponding to $j=1$. The figure consists of one scaling, three global wavelet atoms and twelve detailed wavelet atoms in horizontal, vertical and diagonal direction. 
\subsection{Local Variable Selection in Spatial Point Process}
Suppose $(\Omega,\mathcal{F}, \mathbb{P})$ be a probability space, and the set of observed incidence points, $\left\{\bm{s}_1, \bm{s}_2, \dots, \bm{s}_{n}\right\}$ are irregularly distributed over the spatial region $\mathcal{R}$. Let us assume those incidence points are random realizations of the inhomogeneous spatial Poisson point process $\mathcal N$ on $\mathcal R \subseteq \mathbb{R}^2$. Suppose $\mu=\mathbb{E}[\mathcal{N}(\mathcal{R})]$ is the expected number of points in the spatial region, $\mathcal{R}$, which can be expressed as the integral of the intensity function over the spatial region $\mathcal R$. The volume of the observed spatial region, $\mathcal R$, is defined as $\mathrm{Vol.}(\mathcal R) = \nu$ having a crucial role at the time of defining the likelihood function.

In this section, we discuss the local-level variable selection method for the inhomogeneous Poisson point process via regularized intensity function estimation. Define the predictors as \(\bm X = \left[X_1, \dots, X_{p}\right]\in \R^{p} \) and for the high-dimensional setup we consider $p\propto \mu$. We want to point out the relevant predictors at local level during the estimation of the intensity function. Suppose the intensity function $\pi(\bm s)$ and a second-order product density $\pi_2(\bm s,\bm s')$ exists, then, the pair correlation function is defined by, $\rho(\bm s, \bm s') = \frac{\pi_{2}(\bm s, \bm s')}{\pi(\bm s)\cdot \pi(\bm s')}$, whenever $\pi(\bm s)$ and $\pi(\bm s')$ are positive. This pairwise correlation function measures the model's departure from a Poisson point process. According to Campbell's theorem, for any measurable $\varphi_1:\mathcal R\to[0,\infty)$ and $\varphi_2:\mathcal R\times\mathcal R\to[0,\infty)$,
  \[
  \mathbb E\Big\{\sum_{\bm s\in\mathcal N}\varphi_1(\bm s)\Big\}=\int_{\mathcal R}\varphi_1(\bm s)\pi(\bm s)\,d\bm s,\qquad
  \mathbb E\Big\{\sum_{\bm s_1\ne\bm s_2\in\mathcal N}\varphi_2(\bm s_1,\bm s_2)\Big\}=\int_{\mathcal R\times\mathcal R}\varphi_2(\bm s_1,\bm s_2)\pi_2(\bm s_1,\bm s_2)\,d\bm s_1d\bm s_2.
  \]
 In the literature \cite{thurman2015regularized} selects the predictors via LASSO penalty whereas \cite{choiruddin2023adaptive} has employed adaptive-LASSO (AL) and the Dantzig selector for the same goal. To the best of our knowledge, this is the first work where we will consider the variable selection at a local-level for the localized intensity function. In our study we assume that the intensity function for the inhomogeneous point process depends on $p$ predictors, such that, $\pi(\bm s,\bm \beta) = g^{-1} \left(\beta_0 + \sum_{i=1}^{p} \beta_i(\bm s) X_i(\bm s)\right)$ and \(g(\cdot)\) is the link function. In this Poisson point process, the link function is $g(\cdot) =\log(\cdot)$ and our inhomogeneous intensity function is looking like,
\begin{equation}\label{eq:intensity_surface}
    \pi(\bm s, \bm \beta)  = \exp\left\{\beta_0 + \sum_{i=1}^{p} \beta_i(\bm s) X_i(\bm s)\right\}.
\end{equation}
Suppose the incidence point $\bm u = \left(u_1, u_2\right)^\top$ where $u_1$ and $u_2$ are the coordinates of the observed incidence point respectively. We decompose the coefficeint surface corresponding to the $i$th predictor defined in \eqref{eq:intensity_surface} for a fixed $J$ described in the following.
For $k\in\{0,\dots,2^{J-1}-1\}$ let $I_k=\big[k/2^{J-1},(k+1)/2^{J-1}\big)$, and let $I_k^{(1)}=\big[k/2^{J-1},(k+\tfrac12)/2^{J-1}\big)$ and $I_k^{(2)}=\big[(k+\tfrac12)/2^{J-1},(k+1)/2^{J-1}\big)$ be its left and right halves. Writing $\mathbf{k}=(k_1,k_2)$ and $\sum_{\mathbf{k}}=\sum_{k_1=0}^{2^{J-1}-1}\sum_{k_2=0}^{2^{J-1}-1}$, the coefficient surface of the $i$th predictor is
\begin{equation}
\label{eq:coeff_decompose}
\scriptsize 
\begin{aligned}
\beta_i(\bm u)
=\sum_{\mathbf{k}}2^{J-1}\Big[\,
&\alpha_{i\mathbf{k}}\,\mathbf I\{\bm u\in I_{k_1}\times I_{k_2}\}
+\;\gamma^{(2)}_{i\mathbf{k}}\big(\mathbf I\{\bm u\in I^{(1)}_{k_1}\times I_{k_2}\}-\mathbf I\{\bm u\in I^{(2)}_{k_1}\times I_{k_2}\}\big)\\
+\;&\gamma^{(3)}_{i\mathbf{k}}\big(\mathbf I\{\bm u\in I_{k_1}\times I^{(1)}_{k_2}\}-\mathbf I\{\bm u\in I_{k_1}\times I^{(2)}_{k_2}\}\big)\\
+\;&\gamma^{(4)}_{i\mathbf{k}}\big(\mathbf I\{\bm u\in I^{(1)}_{k_1}\times I^{(1)}_{k_2}\}-\mathbf I\{\bm u\in I^{(1)}_{k_1}\times I^{(2)}_{k_2}\}
-\mathbf I\{\bm u\in I^{(2)}_{k_1}\times I^{(1)}_{k_2}\}+\mathbf I\{\bm u\in I^{(2)}_{k_1}\times I^{(2)}_{k_2}\}\big)\Big].
\end{aligned}
\end{equation}
Since $\phi(2^{J-1}u-k)=\mathbf I\{u\in I_k\}$ and $\psi(2^{J-1}u-k)=\mathbf I\{u\in I^{(1)}_k\}-\mathbf I\{u\in I^{(2)}_k\}$, the coefficient surface in \eqref{eq:coeff_decompose} can be written in terms of the two-dimensional Haar scaling function similar to \eqref{eq:function-decompose} becomes
\begin{equation}
\label{eq:coeff_wavelet}
\beta_i(\bm u)=\sum_{\mathbf{k}}\alpha_{i\mathbf{k}}\,\Phi_{J-1,\mathbf{k}}(\bm u)+\sum_{l=2}^{4}\sum_{\mathbf{k}}\gamma^{(l)}_{i\mathbf{k}}\,\Psi^{(l)}_{J-1,\mathbf{k}}(\bm u),
\end{equation}
where $\Phi_{J-1,\mathbf{k}}(\bm u)$ and $\Psi^{(l)}_{J-1,\mathbf{k}}(\bm u)$ can be obtained by replacing $j =J-1$ in \eqref{eq:tensor_prod}. Here the main idea is we split the entire domain into some disjoint sub-blocks. The expansion of the coefficient surface in \eqref{eq:coeff_decompose} illustrates that the coefficient function is constant over those sub-blocks and it may change its sign alternatively. This idea is not exactly similar to the varying coefficient type modeling in the literature like \cite{coeurjolly2022convolution} and \cite{yin2021fused} because we only consider the finer resolution instead of considering all the resolution levels. Finally this idea bridges the wavelet decomposition and localization of the coefficient surface. For example if $\gamma^{(2)}_{ik_1k_2} \ne 0$ that implies there will exist a non-zero signal within $\left[\frac{k_1}{2^{J-1}}, \frac{k_1+1/2}{2^{J-1}}\right)\times \left[\frac{k_2}{2^{J-1}}, \frac{k_2+1}{2^{J-1}}\right)$ or in  $\left[\frac{k_1+1/2}{2^{J-1}}, \frac{k_1+1}{2^{J-1}}\right)\times \left[\frac{k_2}{2^{J-1}}, \frac{k_2+1}{2^{J-1}}\right)$. Using the two-dimensional wavelet decomposition expression, in \eqref{eq:coeff_decompose} assuming $\delta^{(1)}_{ik_1k_2} = \alpha_{ik_1k_2}, \delta^{(l)}_{ik_1k_2} = \gamma^{(l)}_{ik_1k_2}$ for $l=2,3,4$ and $\tilde{\Psi}^{(1)}_{J-1,k_1k_2} = \Phi_{J-1,k_1, k_2}, \tilde{\Psi}^{(l)}_{J-1,k_1k_2} = \Psi^{(l)}_{J-1,k_1, k_2}$ for $l=2,3,4$, we can write the intensity function of \eqref{eq:intensity_surface} as follows: 
\begin{equation}\label{eq:local-intensity-logit}
\begin{split}
    \pi(\bm s, \bm \delta)= \exp\left\{\beta_0 + \sum_{i=1}^{p}\sum_{l=1}^4\sum_{k_1=0}^{2^{J-1}-1} \sum_{k_2 =0}^{2^{J-1}-1} \delta_{ik_1k_2}^{(l)}\tilde{\Psi}^{(l)}_{J-1,k_1,k_2}(\bm s) X_i(\bm s)\right\}\\
      = \exp\left \{\beta_0 + \sum_{i=1}^{p} \sum_{r(l,k_1,k_2) = 1}^{R}  \delta_{i,r(l,k_1,k_2)} \tilde{\Psi}_{r(l,k_1,k_2)} X_i(\bm s) \right\}.
\end{split}
\end{equation}
In the second line of the expression in \eqref{eq:local-intensity-logit} the total number of wavelet basis at resolution $j=J-1$ is $R = 4\cdot 4^{J-1}$ and the index $r(l,k_1,k_2)\equiv r$ involves different types of wavelet like scaling, horizontal, vertical and diagonal wavelet and also its corresponding shifting parameters, $k_1$ and $k_2$. For notational ease we consider $z_{(i-1)R+r}(\bm s) = \tilde{\Psi}_r(\bm s) X_i(\bm s)$ which is the vectorized version of local design matrix and then for computational ease the inhomogeneous intensity function in \eqref{eq:local-intensity-logit} can be expressed as $ \pi(\bm s, \bm \delta)= \exp\left \{\beta_0 + \sum_{i=1}^{p} \sum_{r = 1}^R  \delta_{(i-1)R + r} z_{(i-1)R+r}(\bm s) \right\} = \exp\{\bm z(\bm s)^\top \, \bm \delta\}$ assuming, $\bm \delta = \left(\delta_0, \delta_1, \dots, \delta_P\right)^\top \in \mathbb{R}^{P+1}, \bm z(\bm s) = \left(1, \bm z_1(\bm s), \dots, \bm z_P(\bm s)\right)$, the intercept term, $\delta_0 = \beta_0$, and $P=p\cdot R$, which is the the number of predictors after vectorization. With the help of Campbell's Theorem, we have 
\[
\begin{aligned}
    \mu &= \int_{\mathcal{R}} \exp\left\{\beta_0 + \sum_{i=1}^{p} \beta_i(\bm s) X_i(\bm s)\right\} d \bm s = \int_{\mathcal{R}} \exp\left \{\beta_0 + \sum_{i=1}^{p} \sum_{r = 1}^R  \delta_{(i-1)R + r} z_{(i-1)R+r}(\bm s) \right\} d\bm s.
\end{aligned}
\]
The logarithm of the composite likelihood function and its first-order derivative become: 
\begin{equation}\label{eq:log-lik-point}
\begin{split}
     \ell(\bm \delta) = \sum_{a=1}^n \bm z_a^\top\, \bm \delta - \int_{\mathcal{R}} \pi(\bm s; \bm \delta) d\bm s,
    \quad \ell^\prime(\bm \delta) = \frac{\partial \ell(\bm \delta)}{\partial \bm \delta}=\sum_{a=1}^n \bm z_a- \int_{\mathcal{R}} \bm z(\bm s)  \pi(\bm s; \bm \delta) d\bm s.
\end{split}
\end{equation}
We have a diverging number of predictors where \(P \propto \mu \) since we have already assumed that $p \propto \mu $. Under the convex LASSO penalty, the loss function, or equivalently the penalized composite likelihood (considering composite-log-likelihood from \eqref{eq:log-lik-point}) for an inhomogeneous Poisson point process, becomes
\begin{equation}\label{eq:lasso-intens}
    Q(\bm \delta) =  \ell(\bm \delta) - \lambda\sum_{i=1}^{P}|\delta_i|, \quad \widehat{\bm \delta} = \arg \max_{\bm \delta \in \R^{P+1}} Q(\bm \delta).
\end{equation}
In (\ref{eq:lasso-intens}) we consider that the penalty parameter, $\lambda$, is a non-negative parameter, and if \(\lambda = 0\) the estimator, \(\widehat{\bm \delta}\) is reduced to the maximum composite likelihood estimator. This estimator simultaneously provides parameter estimates and selects variables at the local level.

We consider another popular non-convex penalty function, SCAD of \cite{Fan01122001}. Under this SCAD penalty, $\mathcal{P}_{\check \lambda}(\theta; \check{\tau})$ with its tuning parameters \(\check{\tau} >2, \check{\lambda} >0\) will be 
\begin{equation}\label{eq:scad-penalty}
   \mathcal{P}_{\check{\lambda}}(\theta;\check{\tau}) =  \begin{cases}
        \check{\lambda} \lvert \theta \rvert; & \lvert \theta \rvert \leq \check{\lambda},\\
        -\frac{\left(\theta^{2} - 2\cdot \check{\tau}\cdot \check{\lambda} \lvert \theta \rvert + \check{\lambda}^{2}\right)}{2(\check{\tau}-1)}; & \check{\lambda} < \lvert \theta \rvert \leq \check{\tau}\cdot \check{\lambda},\\
        \frac{(\check{\tau}+1)\cdot \check{\lambda}^{2}}{2}; & \lvert \theta \rvert > \check{\tau}\cdot \check{\lambda}
    \end{cases}
\end{equation}
and \(\mathcal{P}_{\check{\lambda}}(0;\check{\tau}) = 0\). The first derivative of $\mathcal{P}_{\check{\lambda}}(\theta;\check{\tau})$ is 
\begin{equation}\label{eq:scad-derivative}
\mathcal{P}'_{\check\lambda}(\theta;\check\tau)=\check\lambda\Big\{\mathbf I(\theta\le\check\lambda)
+\frac{(\check\tau\check\lambda-\theta)_+}{(\check\tau-1)\check\lambda}\,\mathbf I(\theta>\check\lambda)\Big\},\quad \theta\ge 0 .
\end{equation}
Under the SCAD penalty in \eqref{eq:scad-penalty} the penalized composite-likelihood becomes
\begin{equation}\label{eq:scad-intens}
    \check{Q}(\check{\bm \delta}) = \ell(\check{\bm \delta}) - \sum_{i=1}^{P}\mathcal{P}_{\check{\lambda}}(|\check{\delta}_i|), \quad \widehat{\check{\bm \delta}} = \arg \max_{\check{\bm \delta} \in \R^{P+1}} \check{Q}(\check{\bm \delta}).
\end{equation}
From these two methods, local LASSO-based intensity function (LLI) estimation in (\ref{eq:lasso-intens}) and local-SCAD-based intensity function estimator (LSI) in (\ref{eq:scad-intens}), we can select the predictors at the local level. Now we will discuss how the estimated sets provide insight into local variable selection. If a predictor is relevant at a location, it has at least one non-zero coefficient attached to some basis function that is non-zero at that location; the converse need not hold, since contributions of several basis functions can cancel. To avoid confusion about the estimated local activation set, we explain with the help of one example.
\begin{example}
Let's fix the maximum resolution level, $J=2$. Then corresponding to the finest resolution level, $j=1$ we have the total $R = 16$ wavelet basis functions. Also, assume we have $p = 50$ predictors; as a result, $P = 800$ and consider $\beta_1(\bm s_1), \beta_2(\bm s_1), X_1(\bm s_1), X_2(\bm s_1)$ are non-zero. Now, the structure of localized design matrix excluding the intercept term, after vectorization, the column index of the localized design matrix becomes 
    \[
\underbrace{\underbrace{1,2,\ldots, 16}_{j=1}}_{X_1},
   \underbrace{\underbrace{17,18,\ldots, 32}_{j=1}}_{X_2}, 
    \underbrace{\underbrace{33,\ldots,48}_{j=1}}_{X_3}, \dots, \underbrace{\underbrace{785,\ldots,800}_{j=1}}_{X_{50}}.
    \]
    For example, the column index starting from $1-16$ arises from the interaction of $X_1$ and wavelets. In this way, it will proceed up to the $50$th predictor. Suppose corresponding to $\bm s_1$, if $\hat{\beta}_1(\bm s_1)$ is non-zero there will exist at-least one coefficient within the set, $\{\hat \delta_1, \dots,\hat \delta_{16}\}$ will be non-zero and for non-zero $\hat{\beta}_2(\bm s_1)$ there will exist at-least one coefficient within the set, $\{\hat \delta_{17}, \dots, \hat \delta_{32}\}$ will be non-zero.
\end{example}
Thus for a particular location $\bm s$, the true set of relevant variables for that specific location is $\mathcal{J}(\bm s) \equiv \{i \in \{1,\ldots, p\}: \beta_i(\bm s) \ne 0\}$, and for that location the locally estimated relevant predictor set for LLI and LSI are denoted respectively by,
\begin{equation}\label{eq:local-vs}
\begin{split}
    &\widehat{\mathcal{J}}(\bm s) \equiv \{i \in \{1,\ldots, p\}: \widehat{\beta}_i(\bm s) \ne 0\},\quad \widehat{\check{\mathcal{J}}}(\bm s) \equiv \{i \in \{1,\ldots, p\}: \widehat{\check{\beta}}_i(\bm s) \ne 0\},
\end{split}
\end{equation}
In \eqref{eq:local-vs} the estimated coefficeint, $\widehat{\beta}_i(\bm s) \ne 0$ implies that there exists at least one $(l^\star, k^\star_1, k^\star_2)$ such that $  \widehat{\delta}^{(l^\star)}_{ik^\star_1k^\star_2}\ne 0$, and $\tilde{\Psi}_{J-1,(k^\star_1, k^\star_2)}^{(l^\star)}(\bm s) \ne 0$. A similar explanation is valid for the coefficient of LSI as well where, $\widehat{\beta}_i(\bm s)$ will be replaced by $\widehat{\check \beta}_i(\bm s)$ and $\widehat{\delta}^{(l^\star)}_{ik^\star_1k^\star_2}$ will be replaced by $\widehat{\check \delta}^{(l^\star)}_{ik^\star_1k^\star_2}$ in \eqref{eq:local-vs}. Thus, using \eqref{eq:lasso-intens} and \eqref{eq:scad-intens}, we can select the location-specific relevant variables. The global activation set for LLI and LSI becomes
\begin{equation}\label{eq:global-vs}
    \widehat{\mathcal A} = \bigcup_{\bm s \in \mathcal{R}}\widehat{\mathcal J}(\bm s),\qquad \widehat{\widetilde{\mathcal A}} = \bigcup_{\bm s \in \mathcal{R}}\widehat{\check{\mathcal J}}(\bm s).
\end{equation}
These sets in \eqref{eq:global-vs} only provides the set of selected variables unlike the location specific information of variable selection as of \eqref{eq:local-vs}. Thus our proposed LLI and LSI satisfy our objective.
\subsection{Computational Details}\label{ss:computation}
In this section, we describe the local variable selection for the Poisson point process by combining the Berman-Turner (BT) approximation (\cite{baddeley2014logistic}) with a multiresolution (Haar) basis to select the predictors at the local level in a high-dimensional setup.  In the domain $\mathcal R$ we have $n$ observed data points which are the realization of the spatial point process, $\mathcal N$ over $\mathcal R$. We first partition the spatial domain $\mathcal R$ into a set of grid-cells of equal area and construct quadrature dummy points at representative locations, such as the pixel centers. The observed event locations and the dummy points are then combined to construct the set of Berman-Turner quadrature points, $\{\bm q_m\}_{m=1}^{N} = \{\bm s_a\}_{a=1}^{n} \cup
\{\bm v_k\}_{k=1}^{n_1}$ where $\bm s_a$ are observed locations, $\bm v_k$ are dummy points, and $N = n+n_1$. Associated with each quadrature point, $\bm q_m$ the quadrature
weight is $\omega_m$ such that, $\sum_{m=1}^N \omega_m = \nu$. We define the indicator variable
\[
\Delta_m =
\begin{cases}
1, & \text{if $\bm q_m$ corresponds to an observed event point},\\
0, & \text{if $\bm q_m$ corresponds to a dummy point}.
\end{cases}
\] 
In the BT quadrature approximation, we approximate the integral term $\int_{\mathcal R} \pi(\bm s, \bm \delta) \, d\bm s$ in \eqref{eq:log-lik-point} as $\int_{\mathcal R} \pi (\bm s, \bm \delta) \, d\bm s\approx \sum_{m=1}^N \omega_m \pi (\bm q_m , \bm \delta)$ with the $N$ points in $\mathcal R$. This quadrature approximation can be easily executed using the \texttt{spatstat} package in R. Altogether, the entire spatial domain, $\mathcal R$ is rescaled and transformed into $[0,1]^2$ for evaluating the two-dimensional Haar-wavelet basis $\tilde{\bm \Psi}$ in the most finer-scale. After evaluating the wavelet basis we construct the $P$-dimensional localized design matrix, $\bm z \in \R^{N \times P}$ likewise \eqref{eq:local-intensity-logit}. Then, after employing the BT quadrature approximation, the log-composite likelihood is
\begin{equation}\label{eq:bt-approx-loglik}
    \ell(\bm \delta) = \sum_{m=1}^N\omega_m \left\{\tilde y_m \bm z_m^\top \bm \delta - \exp\left\{\bm z_m^\top \bm \delta\right\}\right\}, \quad \text{where}\quad \tilde{y}_m = \omega_m^{-1} \Delta_m.
\end{equation}
Next, we employ iteratively reweighted least square (IRLS) to obtain a local quadratic approximation to the smooth BT approximated log-composite likelihood function in \eqref{eq:bt-approx-loglik}. Within each IRLS iteration, the resulting penalized weighted least-squares problem is solved using coordinate descent. Suppose at the current IRLS iteration the coefficient estimate is $\tilde{\bm \delta}$, then the second order Taylor's approximation around $\tilde{\bm \delta}$ is 
\begin{equation}\label{eq:taylor_cli}
    \ell(\bm \delta) \approx \ell(\tilde{\bm \delta}) + \sum_{m=1}^N \omega_m\left(\tilde{y}_m - \exp\left\{\bm z_m^\top \tilde{\bm \delta}\right\}\right)\left(\bm z_m^\top \bm \delta - \bm z_m^\top \tilde{\bm \delta}\right)-\frac{1}{2}\sum_{m=1}^N \omega_m \exp\left\{\bm z_m^\top \tilde{\bm \delta}\right\}\left(\bm z_m^\top \bm \delta - \bm z_m^\top \tilde{\bm \delta}\right)^2.
\end{equation}
Next, we define the working response as $\tilde{y}_m^\ast = \bm z_m^\top \tilde{\bm \delta} + \frac{\tilde{y}_m - \exp\left\{\bm z_m^\top \tilde{\bm \delta}\right\}}{\exp\left\{\bm z_m^\top \tilde{\bm \delta}\right\}}$ and assume $\tilde{\omega}_m = \omega_m \exp\left\{\bm z_m^\top \tilde{\bm \delta}\right\}$ then we can express 
\begin{equation}\label{eq:working-res}
    \begin{split}
\omega_m
\left\{
\widetilde{y}_m -
\exp\!\left(
\bm z_m^{\top}\widetilde{\bm \delta}
\right)
\right\}
=
\omega_m\widetilde{y}_m
-
\widetilde{\omega}_m
=
\frac{
\widetilde{\omega}_m\widetilde{y}_m
}{
\exp\!\left(
\bm z_m^{\top}\widetilde{\bm \delta}
\right)
}
-
\widetilde{\omega}_m
\\
=
\widetilde{\omega}_m
\left[
\frac{
\widetilde{y}_m
}{
\exp\!\left(
\bm z_m^{\top}\widetilde{\bm \delta}
\right)
}
-1
\right]= \widetilde{\omega}_m
\left(
\widetilde{y}_m^{*}
-
\bm z_m^{\top}\widetilde{\bm \delta}
\right).       
    \end{split}
\end{equation}
After replacing \eqref{eq:working-res} in the second-order Taylor expansion in \eqref{eq:taylor_cli}, we obtain
\begin{align*}
\ell(\bm \delta)
\approx\;&
\ell(\widetilde{\bm \delta})+
\sum_{m=1}^{N}
\widetilde{\omega}_m\left(\widetilde{y}_m^{\ast}-\bm z_m^{\top}\widetilde{\bm \delta}\right)
\left(\bm z_m^{\top}\bm \delta-\bm z_m^{\top}\widetilde{\bm \delta}
\right)-
\frac{1}{2}\sum_{m=1}^{N}
\widetilde{\omega}_m
\left(\bm z_m^{\top}\bm \delta-\bm z_m^{\top}\widetilde{\bm \delta}
\right)^2\\
\approx\;&
\ell(\widetilde{\bm \delta})
+\frac{1}{2}
\sum_{m=1}^{N}
\widetilde{\omega}_m
\left(
\widetilde{y}_m^{*}-\bm z_m^{\top}\widetilde{\bm \delta}
\right)^2
-\frac{1}{2}
\sum_{m=1}^{N}\widetilde{\omega}_m\left(
\widetilde{y}_m^{\ast}-\bm z_m^{\top}\bm \delta\right)^2.
\end{align*}
Hence, the quadratic approximation can be rewritten as
\begin{equation}
\ell(\bm \delta)\approx\ell_Q(\bm \delta)=-\frac{1}{2}
\sum_{m=1}^{N}\widetilde{\omega}_m
\left(\widetilde{y}_m^{\ast}-\bm z_m^{\top}\bm \delta
\right)^2+\mathrm{Const.}(\widetilde{\bm \delta}),
\label{eq:quadratic-loglikelihood}
\end{equation}
where, $\mathrm{Const.}(\widetilde{\bm \delta})= \ell(\widetilde{\bm \delta})
+\frac{1}{2}\sum_{m=1}^{N}\widetilde{\omega}_m\left(
\widetilde{y}_m^{\ast}-\bm z_m^{\top}\widetilde{\bm \delta}
\right)^2$ does not depend on the new optimization variable $\bm \delta$. We first construct a decreasing sequence of regularization parameters, $\lambda \in [\underline{\lambda}, \overline{\lambda}]$ where the regularization path, $\Lambda$ is a finite set satisfying $\Lambda \subseteq [\underline{\lambda}, \overline{\lambda}]$. The value $\overline{\lambda}$ is chosen to be the smallest value of $\lambda$ for which all penalized coefficient estimates become zero and it decreases to $\underline{\lambda}$. For each $\lambda$, an outer IRLS loop is used to update $\widetilde{\omega}_m$ and $\widetilde{y}_m^{\ast}$. Within each outer
iteration, coordinate descent is used to solve the penalized weighted least-squares problem. After removing the constant term
$\mathrm{Const.}(\widetilde{\bm \delta})$, the optimization objective in \eqref{eq:quadratic-loglikelihood} reduces to:
\begin{equation*}
\widehat{\bm \delta}_{\lambda}
=\underset{\bm \delta\in\mathbb{R}^{P+1}}{\operatorname{argmin}}\left\{
\frac{1}{2}\sum_{m=1}^{N}
\widetilde{\omega}_m\left(
\widetilde{y}_m^{\ast}-\bm z_m^{\top}\bm \delta
\right)^2+\lambda
\sum_{i=1}^{P}|\delta_i|
\right\}.
\end{equation*}
The coordinate-descent updates for the lasso differ from the ordinary weighted least-squares updates by using the soft-thresholding step; for a detailed description, please see \cite{choiruddin2018convex}. The coordinate descent for LLI can be easily executed using \texttt{glmnet}. For the LSI, the soft-thresholding step differs from the LLI. Unlike the LASSO, the SCAD penalty is non-convex, so the soft-thresholding update no longer applies directly. Although coordinate descent can be applied directly to SCAD, we instead use the local linear approximation (LLA), which locally approximates the SCAD penalty function. The BT approximation in \eqref{eq:bt-approx-loglik} can be written as a weighted Poisson log-likelihood with response $\tilde y_m=\Delta_m/\omega_m$ and prior weights $\omega_m$, or, equivalently up to an additive constant, as a Poisson
log-likelihood with response $\Delta_m$ and offset $\log\omega_m$ because, $\sum_{m}\omega_m\{\tilde y_m\bm z_m^\top\bm\delta-\exp(\bm z_m^\top\bm\delta)\}
=\sum_{m}\{\Delta_m(\bm z_m^\top\bm\delta+\log\omega_m)-\exp(\bm z_m^\top\bm\delta+\log\omega_m)\}
-\sum_m\Delta_m\log\omega_m$.
Either form requires a solver that accepts observation-specific weights or an offset,
which \texttt{ncvreg} does not support (see also \cite{choiruddin2018convex}).
We therefore implement the SCAD fit through the LLA, which turns each step into a
weighted LASSO problem \eqref{eq:lsi-irls-lla} that can be solved with \texttt{glmnet}
using coefficient-specific penalty factors $w_i^{(t)}$. Suppose $\check{\tilde{\bm \delta}}^{(t)}$ denotes the estimated coefficient in $t$-th LLA iteration. The SCAD penalty function from \eqref{eq:scad-penalty} can be approximated using its derivative from \eqref{eq:scad-derivative} around $\check{\tilde{\delta}}_i^{(t)}$ by employing first-order Taylor expansion as follows
\begin{equation*}
\mathcal{P}_{\check{\lambda}}(|\check{\delta}_i|; \check{\tau}) \approx\; \mathcal{P}_{\check \lambda}\left(|\check{\tilde{\delta}}^{(t)}_i|; \check{\tau}\right) + \mathcal{P}_{\check{\lambda}}' \left(|\check{\tilde{\delta}}^{(t)}_i|; \check{\tau}\right)\left(|\check\delta_i|-|\tilde{\check{\delta}}^{(t)}_i|\right).        
\end{equation*}
  Since $\mathcal P_{\check\lambda}(|\check{\tilde\delta}^{(t)}_i|;\check\tau)$ and $\mathcal P^\prime_{\check\lambda}(|\check{\tilde\delta}^{(t)}_i|;\check\tau)\,|\check{\tilde\delta}^{(t)}_i|$ do not depend on $\check\delta_i$, they can be absorbed into a constant. Therefore the penalty function can be approximated as:
\[
\mathcal{P}_{\check{\lambda}}(|\check{\delta}_i|; \check{\tau}) \approx\; \mathcal{P}_{\check{\lambda}}' \left(|\check{\tilde{\delta}}^{(t)}_i|; \check{\tau}\right)|\check\delta_i|+ \text{Const.}
\]
Accordingly, at the $t$th LLA iteration we define the coefficient-specific
penalty factor, $w_i^{(t)} = \mathcal{P}_{\check{\lambda}}' \left(|\check{\tilde{\delta}}^{(t)}_i|; \check{\tau}\right)/\check{\lambda}$ where $i=1,\dots,P$. As a result the LLA approximated SCAD penalty becomes, $\sum_{i=1}^{P} \mathcal{P}_{\check{\lambda}}(|\check{\delta}_i|; \check{\tau}) \approx
\check{\lambda} \sum_{i=1}^{P} w_i^{(t)} |\check{\delta}_i| + \mathrm{Const}$. This representation also provides an intuitive interpretation of the SCAD penalty: if $|\check{\delta}^{(t)}_i|\leq\check{\lambda}$, then $w_i^{(t)}=1$ and the coefficient receives the same amount of penalization as under the LASSO penalty, if $\check{\lambda} < |\check{\delta}^{(t)}_i|\leq\check{\tau}\cdot \check{\lambda}$ then $0<w_i^{(t)}<1$, so that the amount of shrinkage gradually decreases as the magnitude of the coefficient increases. Finally, if $|\check{\delta}^{(t)}_i|>\check{\tau}\cdot \check{\lambda}$, then
$w_i^{(t)}=0$, and as a result the sufficiently large coefficient is left unpenalized. This is the key distinction between the SCAD and LASSO penalties where SCAD reduces the shrinkage bias associated
with large nonzero coefficients. Combining the LLA approximation with the IRLS quadratic approximation in \eqref{eq:quadratic-loglikelihood}, the
$(t+1)$st LSI update is obtained from
\begin{equation}\label{eq:lsi-irls-lla}
\widehat{\check{\bm\delta}}_{\check{\lambda}}^{(t+1)}
= \underset{\check{\bm\delta}\in\mathbb R^{P+1}}{\operatorname{argmin}} \left\{
\frac{1}{2} \sum_{m=1}^{N} \widetilde{\omega}_m
\left( \widetilde y_m^\ast - \bm z_m^\top\check{\bm\delta} \right)^2 +
\check\lambda \sum_{i=1}^{P}
w_i^{(t)} |\check{\delta}_i| \right\}.
\end{equation}
Although the original SCAD-penalized objective is nonconvex, its LLA
representation is repeatedly solved until
convergence.

Next we select the tuning parameter based on composite BIC (CBIC) score like \cite{choiruddin2021information}, motivated by the generalized information criterion (GIC) of \cite{zhang2010regularization}, and we choose the tuning parameter as follows:
\begin{equation}\label{eq:bic-bt}
\begin{split}
      \mathrm{CBIC}(\lambda)
  \;=\; -2\,\ell(\hat{\bm \delta}_\lambda)
        \;+\; \mathcal K_0(\lambda)\,\log\!\big(\mu\big),\\
     \lambda^\ast \in \arg\min_{\lambda}\ \mathrm{CBIC}(\lambda),
  \qquad
  \widehat{\bm \delta} \;=\; \widehat{\bm \delta}_{\lambda^\ast}.  
\end{split}
\end{equation}
In \eqref{eq:bic-bt} the cardinality of the set of non-zero coefficients, $\hat{\delta}_{\lambda,i}$ corresponding to penalty parameter, $\lambda$ is, $\mathcal{K}_0(\lambda) = \# \left\{i: \hat{\delta}_{\lambda,i} \ne 0\right\}$. This CBIC-based tuning parameter selection actually helps us to select the variables efficiently at the local level. In practice, $\mu$ is unknown therefore we follow \cite{choiruddin2021information}'s suggestion and replace $\mu$ by $\mathcal{N}(\mathcal R)$ in \eqref{eq:bic-bt}. In Algorithm~\ref{alg:bt-haar-local-selection} we describe the workflow of LLI and LSI for spatial point process in detail.

\begin{algorithm}[!t]
\caption{LLI and LSI Estimation for Spatial Poisson Point Processes}
\label{alg:bt-haar-local-selection}
\begin{algorithmic}[1]

\Require Event locations $\{\bm s_a\}_{a=1}^{n}$, covariates
$\{X_i\}_{i=1}^{p}$, spatial domain $\mathcal R$, resolution $J$,
and regularization path $\Lambda$.

\State Construct the Berman--Turner quadrature
$\{\bm q_m,\omega_m,\Delta_m\}_{m=1}^{N}$ and set
$\tilde y_m=\omega_m^{-1}\Delta_m$.

\State Transform $\bm q_m\in\mathcal R$ to $[0,1]^2$ and evaluate Wavelet Basis at the finest resolution
$j=J-1$, giving $\widetilde{\bm\Psi} \in\mathbb R^{N\times R}$, $R=4\cdot4^{J-1}$, and construct the localized design matrix.

\State For each $\lambda\in\Lambda$, update the IRLS weights $\widetilde\omega_m$ and working responses $\widetilde y_m^\ast$, and obtain the LLI estimator by solving the LASSO-penalized weighted least-squares problem in \eqref{eq:lasso-intens}.

\State For each $\check\lambda\in\Lambda$, update the SCAD weights $w_i^{(t)}$ and obtain the LSI estimator by iterating the LLA-IRLS in \eqref{eq:lsi-irls-lla} until convergence.

\State Compute $\mathrm{CBIC}(\lambda)$ along each regularization path and select the tuning parameter minimizing \eqref{eq:bic-bt} for LLI and LSI separately.

\State Using the nonzero localized coefficients, construct the location-specific active predictor sets $\widehat{\mathcal J}(\bm s)$ and $\widehat{\widetilde{\mathcal J}}(\bm s)$
as defined in \eqref{eq:local-vs}.

\State Compute the fitted intensity surface
$\widehat{\pi}(\bm s)$ using the selected localized coefficients.

\end{algorithmic}
\end{algorithm}

\section{Simulation Studies}\label{s:sim}
In this section, we compare the performance of LLI and LSI estimator with global methods: LASSO, SCAD, and adaptive-LASSO-based intensity estimator with increasing $\mu$. In this simulation study we generate the data from Thomas point process and inhomogeneous Poisson point process. We fix the target intensity $\pi_0(\bm u)$ for the spatial window $\mathcal{R}$, and the parents follow a homogeneous Poisson point process with intensity $\kappa$. Parents are generated from a homogeneous Poisson process with intensity $\kappa$, and each parent independently receives a Poisson number of offspring with mean $\pi_{\max}/\kappa$, where $\pi_{\max}=\max_{\bm u\in\mathcal R}\pi_0(\bm u)$. The offspring of a parent at $\bm u^\star$ are placed at $\bm u^\star+\bm\varepsilon$, with $\bm\varepsilon\sim N_2(\bm 0,\sigma^2\bm I_2)$ independently across offspring. Each offspring at $\bm y$ is then retained with probability $\pi_0(\bm y)/\pi_{\max}$, and offspring outside $\mathcal R$ are discarded. The resulting process has intensity $\pi_0(\bm u)=\exp\{\beta_0+\eta_0(\bm u)\}$, with $\eta_0(\bm u)=\sum_{i=1}^p\beta_i(\bm u)X_i(\bm u)$, so the covariates enter through $\pi_0$ while $\kappa$ and $\sigma$ determine the clustering. The resulting process has pair correlation function, $\rho (r)=1+\frac{1}{4\pi\kappa\sigma^2}\exp\Big(-\frac{r^2}{4\sigma^2}\Big)$. The point process is simulated with \texttt{rThomas} in \textsf{spatstat} package, which simulates parents on an expanded window to avoid edge effects. In each scenario the intercept $\beta_0$ is defined as $\beta_0=\log\mu-\log\int_{\mathcal{R}}\exp\{\eta_0(\bm u)\}\,\mathrm d\bm u$. In this simulation framework we consider three scenarios. In Scenario~I we consider $\kappa = 5\times 10^{-4}$ and $\sigma = 20$ and in Scenario~II we generate spatial points from inhomogeneous poisson process whereas in Scenario~III we fix $\kappa = 25$ and $\sigma = 0.0283$. In the Scenario~I the quantity $\rho(0) - 1 = 0.4$ which indicates weaker clustering similar to \cite{choiruddin2018convex}. However in Scenario~III the quantity $\rho(0) - 1 = 3.98$ which indicates strong clustering. 
\begin{figure}
    \centering
    \includegraphics[width=\linewidth]{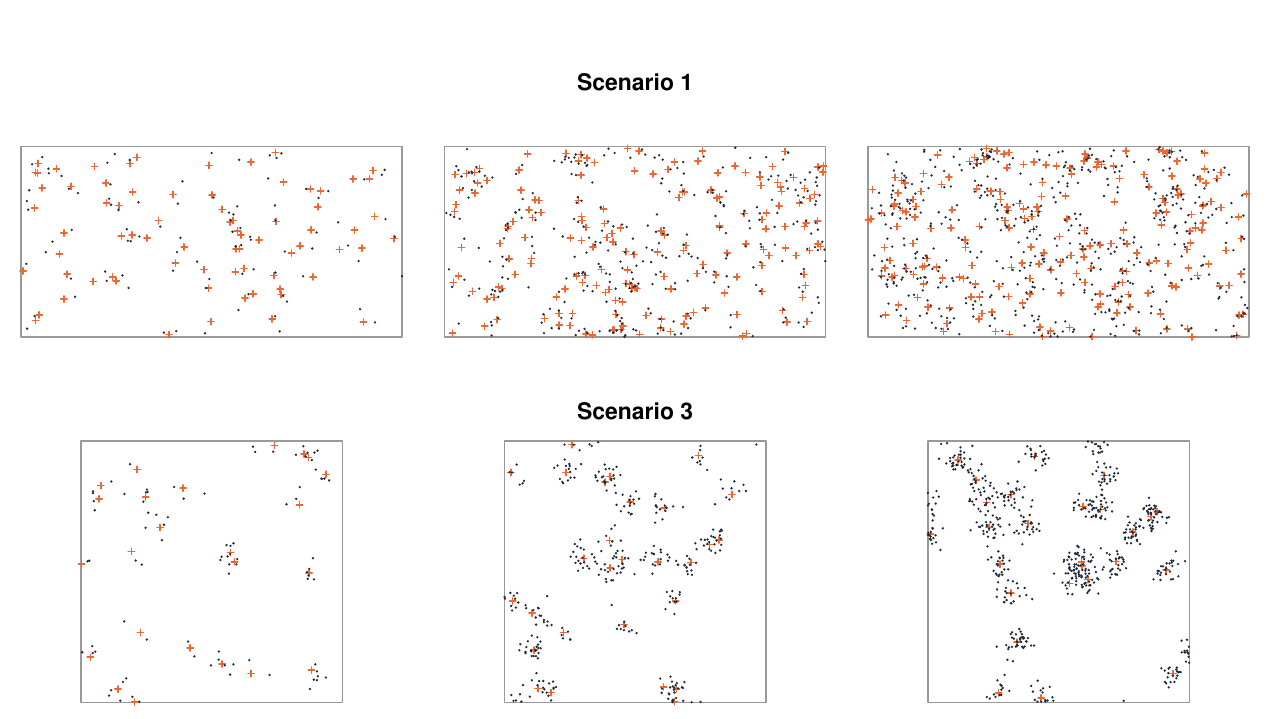}
    \caption{Realisations of the Thomas processes used in the simulation study, for $\mu=100$, $300$ and $600$ (left to right); black dots are points and orange crosses are parents, and the same parents are used within each row. Top, Scenario~I: $\mathcal{R}=[0,1000]\times[0,500]$m, with 0.4, 1.2 and 2.4 points per parent. Most of the parents have at most two offspring, and the pattern is close to complete spatial randomness. Bottom, Scenario~III: the window is $\mathcal{R}=[0,1]^2$, with 4, 12 and 24 points per parent.}
    \label{fig:thomas_pp}
\end{figure}
From Figure~\ref{fig:thomas_pp} we detect that in Scenario~I most parents produce at most two offspring, and the pattern is close to complete spatial randomness. In Scenario~III we consider the spatial window is $\mathcal{R}=[0,1]^2$ on a $16\times16$ grid. In Scenario~III the points are concentrated in about $25$ dense clusters with large empty regions in between. Increasing $\mu$ more points are added to the existing clusters, and makes the cluster denser (from Figure~\ref{fig:thomas_pp}). 

In Scenario~I similar to \cite{choiruddin2018convex} we consider the Barro Colorado Island (BCI) domain  where $\mathcal{R}=[0,1000]\times[0,500]$\,m, is divided into $201\times101$ pixel images. Suppose $\tilde{X}_1$ and $\tilde{X}_2$ be the standardized elevation and gradient images (\texttt{bei.extra} in \textsf{spatstat}) and $\tilde{X}_3,\dots,\tilde{X}_{20}$ independent Gaussian images with mean zero and standard deviation $10^{-1}$. The covariates are $\bm X(\bm u)=\bm V^\top\tilde{\bm X}(\bm u)$ with $\bm V^\top\bm V=\bm\Omega$, $\Omega_{ij}=0.7^{|i-j|}$ and $\Omega_{12}=\Omega_{21}=0$, so that $X_1$ and $X_2$ are elevation and gradient and $p=20$. With $\bm u=(u_1,u_2)$ the spatial window is rescaled to $[0,1]^2$, the coefficient functions corresponding to the first two relevant predictors are
\begin{equation*}
\beta_1(\bm u)=2\,\mathbf I(u_1<0.5,\ u_2<0.5),\, \beta_2(\bm u)=0.75\,\mathbf I(u_1\ge0.5,\ u_2\ge0.5),\,
\beta_3=\dots=\beta_{20}=0
\end{equation*}
As a result the elevation acts only in the lower-left quadrant and gradient only in the upper-right quadrant. The resulting intensity becomes, $\pi_{01}(\bm u)=\exp\{\beta_0+\beta_1(\bm u)X_1(\bm u)+\beta_2(\bm u)X_2(\bm u)\}$ with $\beta_0=\log\mu-\log\left[\int_{\mathcal{R}}\exp\{\beta_1(\bm u)z_1(\bm u)+\beta_2(\bm u)z_2(\bm u)\}\,\mathrm d\bm u\right]$, and this integral is calculated over, $201\times101$ raster with the pixel area $\approx 25$\,m$^2$. Scenario~II is replicated under similar setup only the difference is there spatial points are coming inhomogeneous Poisson point process with intensity $\pi_{01}(\bm u)$. In this simulation study we fix $J=2$ where we consider total $16$ wavelet basis functions.

In Scenario~III there are $p=500$ predictors where, $X_1,\dots,X_{20}$ are Gaussian random fields generated from the exponential covariance function, with unit variance and range $0.25$. After rescaling predictor satisfies, $1\le X_i \le 2$ for $i=1,\dots, 20$ and $X_i \overset{\mathrm{iid}}{\sim} N(0,0.001)$ for $21\le i \le 500$. The coefficient functions of the first ten predictors as follows:
\[
  \beta_1(\bm{u})
=
\frac{1}{2}
\left[
\mathbf{I}\left\{0\leq u_1<\frac{1}{4}\right\}
-
\mathbf{I}\left\{\frac{1}{4}\leq u_1<\frac{1}{2}\right\}
\right]
\mathbf{I}\left\{0\leq u_2<\frac{1}{2}\right\},
\]
\[
\beta_2(\bm{u})
=
\frac{1}{2}
\left[
\mathbf{I}\left\{0\leq u_1<\frac{1}{4}\right\}
-
\mathbf{I}\left\{\frac{1}{4}\leq u_1<\frac{1}{2}\right\}
\right]
\mathbf{I}\left\{\frac{1}{2}\leq u_2\leq 1\right\},
\]
\[
\begin{aligned}
\beta_3(\bm{u})
&=
\frac{1}{2}
\left[
\mathbf{I}\left\{\frac{1}{2}\leq u_1<\frac{3}{4}\right\}
-
\mathbf{I}\left\{\frac{3}{4}\leq u_1\leq 1\right\}
\right]
\mathbf{I}\left\{0\leq u_2<\frac{1}{2}\right\},
\end{aligned}
\]
\[
\begin{aligned}
 \beta_4(\bm{u})
&=
\frac{1}{2}
\left[
\mathbf{I}\left\{\frac{1}{2}\leq u_1<\frac{3}{4}\right\}
-
\mathbf{I}\left\{\frac{3}{4}\leq u_1\leq 1\right\}
\right]
\mathbf{I}\left\{\frac{1}{2}\leq u_2\leq 1\right\},
\\[2mm]
\beta_5(\bm{s})
&=
\frac{1}{2}
\mathbf{I}\left\{0\leq u_1<\frac{1}{2}\right\}
\left[
\mathbf{I}\left\{0\leq u_2<\frac{1}{4}\right\}
-
\mathbf{I}\left\{\frac{1}{4}\leq u_2<\frac{1}{2}\right\}
\right],   
\end{aligned}
\]
\[
\begin{aligned}
  \beta_6(\bm{u})
&=
\frac{1}{2}
\mathbf{I}\left\{0\leq u_1<\frac{1}{2}\right\}
\left[
\mathbf{I}\left\{\frac{1}{2}\leq u_2<\frac{3}{4}\right\}
-
\mathbf{I}\left\{\frac{3}{4}\leq u_2\leq 1\right\}
\right],
\\[2mm]
\beta_7(\bm{u})
&=
\frac{1}{2}
\mathbf{I}\left\{\frac{1}{2}\leq u_1\leq 1\right\}
\left[
\mathbf{I}\left\{0\leq u_2<\frac{1}{4}\right\}
-
\mathbf{I}\left\{\frac{1}{4}\leq u_2<\frac{1}{2}\right\}
\right],  
\end{aligned}
\]
\begin{equation*}
\begin{split}
\beta_8(\bm{u})
&=
\frac{1}{2}
\mathbf{I}\left\{\frac{1}{2}\leq u_1\leq 1\right\}
\left[
\mathbf{I}\left\{\frac{1}{2}\leq u_2<\frac{3}{4}\right\}
-
\mathbf{I}\left\{\frac{3}{4}\leq u_2\leq 1\right\}
\right],
\\[2mm]
\beta_9(\bm{u})
&=
\frac{1}{2}
\left[
\mathbf{I}\left\{0\leq u_1<\frac{1}{4}\right\}
-
\mathbf{I}\left\{\frac{1}{4}\leq u_1<\frac{1}{2}\right\}
\right]
\left[
\mathbf{I}\left\{0\leq u_2<\frac{1}{4}\right\}
-
\mathbf{I}\left\{\frac{1}{4}\leq u_2<\frac{1}{2}\right\}
\right],
\\[2mm]
\beta_{10}(\bm{u})
&=
\frac{1}{2}
\left[
\mathbf{I}\left\{\frac{1}{2}\leq u_1<\frac{3}{4}\right\}
-
\mathbf{I}\left\{\frac{3}{4}\leq u_1\leq 1\right\}
\right]
\left[
\mathbf{I}\left\{\frac{1}{2}\leq u_2<\frac{3}{4}\right\}
-
\mathbf{I}\left\{\frac{3}{4}\leq u_2\leq 1\right\}
\right].
\end{split}
\end{equation*}
The resulting intensity becomes, $\pi_{03}(\bm u)=\exp\{\beta_0+ \sum_{i=1}^{10}\beta_i(\bm u)X_i(\bm u)\}$.

The selection and estimation performance of each method is measured by the true-positive rate (TPR), the false-positive rate (FPR) and the root-mean-square prediction error (RMSPE) of the coefficient surfaces,
\begin{equation}\label{eq:sel-metrics}
\operatorname{TPR}=\frac{[[\mathcal A\cap\widehat{\mathcal A}]]}{[[\mathcal A]]},\qquad
\operatorname{FPR}=\frac{[[\widehat{\mathcal A}\setminus\mathcal A]]}{p-[[\mathcal A]]},\qquad
\operatorname{RMSPE}(\bm\beta)=\Bigg[\frac{1}{p\,\nu}\sum_{i=1}^{p}\int_{\mathcal R}\big\{\widehat\beta_i(\bm u)-\beta_i(\bm u)\big\}^2\,\mathrm d\bm u\Bigg]^{1/2},
\end{equation}
where $\mathcal A$ is the true active set, $\widehat{\mathcal A}$ the set of predictors with at least one non-zero estimated coefficient \eqref{eq:global-vs}, $[[\cdot]]$ denotes cardinality of a set. A higher TPR in \eqref{eq:sel-metrics} means that more truly relevant predictors are selected, a lower FPR that fewer irrelevant predictors are selected, and a lower $\operatorname{RMSPE}(\bm\beta)$ that the coefficient surfaces are estimated more accurately. For LLI and LSI we also report the local TPR and FPR at the observed events, based on the location-specific sets $\mathcal J(\bm s_a)$ and $\widehat{\mathcal J}(\bm s_a)$ in \eqref{eq:local-vs}. For an observed event $\bm s_a$, $a=1,\dots,n$, let $\mathcal J(\bm s_a)=\{i:\beta_i(\bm s_a)\neq0\}$ be the set of predictors that are truly active at $\bm s_a$ and $\widehat{\mathcal J}(\bm s_a)=\{i:\widehat\beta_i(\bm s_a)\neq0\}$ the set selected there \eqref{eq:local-vs}. The local true- and false-positive rates at $\bm s_a$ are
\[
\operatorname{TPR}(\bm s_a)=\frac{[[\mathcal J(\bm s_a)\cap\widehat{\mathcal J}(\bm s_a)]]}{[[\mathcal J(\bm s_a)]]},\qquad
\operatorname{FPR}(\bm s_a)=\frac{[[\widehat{\mathcal J}(\bm s_a)\setminus\mathcal J(\bm s_a)]]}{p-[[\mathcal J(\bm s_a)]]},
\]
and we report their averages over the observed events,
\begin{equation}\label{eq:local-rates}
\overline{\operatorname{TPR}}_{\mathrm{loc}}=\frac{1}{[[\mathcal S_+]]}\sum_{a\in\mathcal S_+}\operatorname{TPR}(\bm s_a),\qquad
\overline{\operatorname{FPR}}_{\mathrm{loc}}=\frac{1}{n}\sum_{a=1}^{n}\operatorname{FPR}(\bm s_a),
\end{equation}
where $\mathcal S_+=\{a:[[\mathcal J(\bm s_a)]]>0\}$ is the set of events at which at least one predictor is truly active. $\overline{\operatorname{TPR}}_{\mathrm{loc}}$ in \eqref{eq:local-rates} is the average proportion of the locally relevant predictors that are recovered at the event locations, and $\overline{\operatorname{FPR}}_{\mathrm{loc}}$ is the average proportion of the locally irrelevant predictors that are wrongly selected there. 

\begin{table}[!htbp]
\centering
\caption{Scenario~I (weak clustering, BCI window, $\kappa=5\times10^{-4}\,\mathrm{m}^{-2}$, $\sigma=20$\,m): mean (standard deviation) over $1000$ replications of TPR, FPR, average local TPR and FPR, RMSPE of the coefficient surfaces, selected penalty, and run time.}
\label{tab:sim-s1}
\resizebox{\linewidth}{!}{%
\begin{tabular}{llrrrrrrr}
\toprule
$\mu$ & Method & TPR & FPR & RMSPE$(\bm \beta)$ 
& $\overline{\operatorname{TPR}}_{\mathrm{loc}}$ & $\overline{\operatorname{FPR}}_{\mathrm{loc}}$ & 
$\lambda$ & Runtime (s)\\
\midrule
\multirow{5}{*}{100}
& LLI  & 0.615 (0.292) & 0.006 (0.017) & 0.216 (0.024) & 0.727 (0.250) & 0.012 (0.019) & 0.0205 (0.0068) & 0.02 (0.00) \\
& LSI  & 0.795 (0.267) & 0.012 (0.023) & 0.179 (0.073) & 0.845 (0.204) & 0.023 (0.021) & 0.0204 (0.0095) & 0.79 (5.08) \\
& GLI  & 0.260 (0.261) & 0.018 (0.026) & 0.243 (0.011) & --            & --            & 0.0080 (0.0062) & 0.01 (0.00) \\
& GSI  & 0.270 (0.260) & 0.015 (0.025) & 0.244 (0.013) & --            & --            & 0.0136 (0.0080) & 0.02 (0.00) \\
& GALI & 0.150 (0.251) & 0.029 (0.028) & 0.253 (0.025) & --            & --            & --              & 0.01 (0.00) \\
\midrule
\multirow{5}{*}{200}
& LLI  & 0.900 (0.201) & 0.016 (0.025) & 0.186 (0.034) & 0.930 (0.128) & 0.029 (0.024) & 0.0216 (0.0085) & 0.02 (0.00) \\
& LSI  & 0.985 (0.086) & 0.016 (0.025) & 0.091 (0.059) & 0.976 (0.113) & 0.029 (0.019) & 0.0210 (0.0047) & 0.22 (0.30) \\
& GLI  & 0.355 (0.228) & 0.024 (0.028) & 0.241 (0.009) & --            & --            & 0.0086 (0.0081) & 0.01 (0.00) \\
& GSI  & 0.320 (0.241) & 0.017 (0.026) & 0.241 (0.011) & --            & --            & 0.0196 (0.0109) & 0.02 (0.00) \\
& GALI & 0.170 (0.238) & 0.038 (0.026) & 0.254 (0.032) & --            & --            & --              & 0.01 (0.00) \\
\midrule
\multirow{5}{*}{300}
& LLI  & 0.965 (0.128) & 0.021 (0.027) & 0.156 (0.037) & 0.975 (0.085) & 0.041 (0.025) & 0.0144 (0.0056) & 0.03 (0.00) \\
& LSI  & 1.000 (0.000) & 0.013 (0.023) & 0.078 (0.038) & 1.000 (0.000) & 0.026 (0.019) & 0.0195 (0.0045) & 0.22 (0.15) \\
& GLI  & 0.405 (0.221) & 0.028 (0.028) & 0.241 (0.008) & --            & --            & 0.0056 (0.0060) & 0.01 (0.00) \\
& GSI  & 0.375 (0.240) & 0.017 (0.026) & 0.240 (0.009) & --            & --            & 0.0189 (0.0114) & 0.03 (0.00) \\
& GALI & 0.280 (0.278) & 0.044 (0.023) & 0.257 (0.041) & --            & --            & --              & 0.01 (0.00) \\
\midrule
\multirow{5}{*}{500}
& LLI  & 1.000 (0.000) & 0.027 (0.028) & 0.139 (0.030) & 0.997 (0.026) & 0.050 (0.028) & 0.0111 (0.0039) & 0.04 (0.00) \\
& LSI  & 1.000 (0.000) & 0.013 (0.024) & 0.063 (0.034) & 1.000 (0.000) & 0.023 (0.022) & 0.0186 (0.0052) & 0.50 (1.09) \\
& GLI  & 0.425 (0.179) & 0.030 (0.028) & 0.240 (0.007) & --            & --            & 0.0053 (0.0052) & 0.01 (0.00) \\
& GSI  & 0.405 (0.197) & 0.024 (0.028) & 0.240 (0.008) & --            & --            & 0.0156 (0.0113) & 0.03 (0.00) \\
& GALI & 0.330 (0.268) & 0.043 (0.023) & 0.250 (0.032) & --            & --            & --              & 0.02 (0.00) \\
\midrule
\multirow{5}{*}{600}
& LLI  & 1.000 (0.000) & 0.030 (0.028) & 0.131 (0.026) & 1.000 (0.000) & 0.053 (0.023) & 0.0113 (0.0036) & 0.04 (0.01) \\
& LSI  & 1.000 (0.000) & 0.014 (0.024) & 0.059 (0.031) & 1.000 (0.000) & 0.024 (0.021) & 0.0191 (0.0050) & 0.48 (0.56) \\
& GLI  & 0.460 (0.154) & 0.033 (0.027) & 0.241 (0.007) & --            & --            & 0.0041 (0.0049) & 0.01 (0.00) \\
& GSI  & 0.410 (0.193) & 0.026 (0.028) & 0.241 (0.009) & --            & --            & 0.0153 (0.0124) & 0.03 (0.00) \\
& GALI & 0.305 (0.245) & 0.046 (0.021) & 0.250 (0.023) & --            & --            & --              & 0.02 (0.00) \\
\bottomrule
\end{tabular}}
\end{table}

Table~\ref{tab:sim-s1} reports the results for Scenario~I. The advantage of the local estimators is clearly visible in this setting. The TPR increases from 61.5\% (LLI) and 79.5\% (LSI) at $\mu=100$ to 100\% for $\mu\ge300$ (LSI) and $\mu\ge500$ (LLI). In contrast, the global estimators recover at most 46\% of the relevant predictors (GLI 26.0--46.0\%, GSI 27.0--41.0\%, GALI 15.0--33.0\%), because a single constant coefficient cannot represent a covariate that acts in only one quadrant. LSI has the lowest RMSPE$(\bm\beta)$ for every $\mu$, decreasing from 0.179 to 0.059, followed by LLI (0.216 to 0.131), whereas the RMSPE of the global estimators stays at about 0.24--0.26 regardless of $\mu$. The local TPR follows the same pattern as the global TPR: the average local TPR is 72.7\% (LLI) and 84.5\% (LSI) at $\mu=100$ and reaches 100\% for $\mu\ge300$ (LSI) and $\mu=600$ (LLI), with LLI already at 99.7\% for $\mu=500$. The local FPR is at most 5.3\% for LLI and 2.9\% for LSI. The global FPR is below 5\% for all methods; it is at most 1.6\% for LSI and is highest for GALI at every $\mu$. The selected $\lambda$ tends to decrease with $\mu$ for GLI (from 0.0080 at $\mu=100$ to 0.0041 at $\mu=600$) and LLI (from 0.0205 to 0.0113), although for both the largest value occurs at $\mu=200$. For GSI the selected $\lambda$ varies between 0.0136 and 0.0196 without a monotone trend, peaking at $\mu=200$, and for LSI it is nearly constant, between 0.0186 and 0.0210, for all $\mu$. We observe that LLI selects a larger $\lambda$ than GLI for every $\mu$, about two to three times as large, which reflects the larger number of columns in the localized design. All methods take less than one second per fit on average. LLI, GLI, GSI and GALI each require at most 0.04 seconds, while LSI is the most expensive (0.22--0.79 seconds); its large standard deviation at $\mu=100$ (5.08 seconds) indicates that a small number of fits took considerably longer than the rest.

\begin{table}[!htbp]
\centering
\caption{Scenario~II (inhomogeneous Poisson process, BCI window, same intensity as in Scenario~I): mean (standard deviation) over $1000$ replications of TPR, FPR, average local TPR and FPR, RMSPE of the coefficient surfaces, selected penalty, and run time.}
\label{tab:sim-s2}
\resizebox{\linewidth}{!}{%
\begin{tabular}{llrrrrrrr}
\toprule
$\mu$ & Method & TPR & FPR & RMSPE$(\bm \beta)$ 
& $\overline{\operatorname{TPR}}_{\mathrm{loc}}$ & $\overline{\operatorname{FPR}}_{\mathrm{loc}}$ & 
$\lambda$ & Runtime (s)\\
\midrule
\multirow{5}{*}{100}
& LLI  & 0.630 (0.262) & 0.004 (0.014) & 0.214 (0.024) & 0.771 (0.203) & 0.007 (0.014) & 0.0199 (0.0059) & 0.02 (0.00) \\
& LSI  & 0.875 (0.218) & 0.005 (0.016) & 0.139 (0.094) & 0.914 (0.141) & 0.017 (0.018) & 0.0199 (0.0109) & 0.20 (0.20) \\
& GLI  & 0.265 (0.251) & 0.014 (0.024) & 0.241 (0.011) & --            & --            & 0.0084 (0.0060) & 0.01 (0.00) \\
& GSI  & 0.280 (0.249) & 0.013 (0.023) & 0.242 (0.013) & --            & --            & 0.0134 (0.0066) & 0.02 (0.00) \\
& GALI & 0.155 (0.232) & 0.030 (0.028) & 0.258 (0.051) & --            & --            & --              & 0.01 (0.00) \\
\midrule
\multirow{5}{*}{200}
& LLI  & 0.925 (0.179) & 0.014 (0.024) & 0.173 (0.032) & 0.958 (0.102) & 0.019 (0.021) & 0.0197 (0.0072) & 0.02 (0.00) \\
& LSI  & 1.000 (0.000) & 0.006 (0.017) & 0.065 (0.044) & 1.000 (0.000) & 0.013 (0.014) & 0.0223 (0.0041) & 0.13 (0.08) \\
& GLI  & 0.455 (0.144) & 0.020 (0.027) & 0.239 (0.008) & --            & --            & 0.0083 (0.0072) & 0.01 (0.00) \\
& GSI  & 0.435 (0.169) & 0.011 (0.022) & 0.238 (0.008) & --            & --            & 0.0206 (0.0091) & 0.02 (0.00) \\
& GALI & 0.190 (0.244) & 0.044 (0.023) & 0.252 (0.022) & --            & --            & --              & 0.01 (0.00) \\
\midrule
\multirow{5}{*}{300}
& LLI  & 0.990 (0.070) & 0.012 (0.023) & 0.154 (0.030) & 0.995 (0.035) & 0.023 (0.021) & 0.0148 (0.0048) & 0.02 (0.00) \\
& LSI  & 1.000 (0.000) & 0.006 (0.017) & 0.050 (0.026) & 1.000 (0.000) & 0.011 (0.013) & 0.0202 (0.0043) & 0.18 (0.06) \\
& GLI  & 0.480 (0.098) & 0.031 (0.028) & 0.240 (0.006) & --            & --            & 0.0048 (0.0058) & 0.01 (0.00) \\
& GSI  & 0.465 (0.128) & 0.011 (0.022) & 0.238 (0.007) & --            & --            & 0.0210 (0.0104) & 0.03 (0.00) \\
& GALI & 0.310 (0.244) & 0.044 (0.022) & 0.253 (0.033) & --            & --            & --              & 0.01 (0.00) \\
\midrule
\multirow{5}{*}{500}
& LLI  & 1.000 (0.000) & 0.009 (0.021) & 0.135 (0.024) & 1.000 (0.000) & 0.021 (0.018) & 0.0119 (0.0029) & 0.03 (0.00) \\
& LSI  & 1.000 (0.000) & 0.000 (0.000) & 0.039 (0.027) & 1.000 (0.000) & 0.003 (0.007) & 0.0206 (0.0050) & 0.23 (0.08) \\
& GLI  & 0.490 (0.070) & 0.041 (0.025) & 0.241 (0.006) & --            & --            & 0.0033 (0.0048) & 0.01 (0.00) \\
& GSI  & 0.465 (0.128) & 0.031 (0.028) & 0.240 (0.007) & --            & --            & 0.0129 (0.0109) & 0.03 (0.00) \\
& GALI & 0.280 (0.269) & 0.048 (0.019) & 0.253 (0.026) & --            & --            & --              & 0.02 (0.00) \\
\midrule
\multirow{5}{*}{600}
& LLI  & 1.000 (0.000) & 0.009 (0.020) & 0.127 (0.024) & 1.000 (0.000) & 0.022 (0.019) & 0.0123 (0.0032) & 0.04 (0.01) \\
& LSI  & 1.000 (0.000) & 0.001 (0.006) & 0.030 (0.020) & 1.000 (0.000) & 0.002 (0.006) & 0.0232 (0.0046) & 0.24 (0.06) \\
& GLI  & 0.500 (0.000) & 0.041 (0.024) & 0.240 (0.005) & --            & --            & 0.0030 (0.0046) & 0.01 (0.00) \\
& GSI  & 0.475 (0.110) & 0.031 (0.028) & 0.240 (0.006) & --            & --            & 0.0131 (0.0119) & 0.03 (0.00) \\
& GALI & 0.310 (0.254) & 0.051 (0.015) & 0.261 (0.044) & --            & --            & --              & 0.02 (0.00) \\
\bottomrule
\end{tabular}}
\end{table}
Table~\ref{tab:sim-s2} reports the results when the points are generated from an inhomogeneous Poisson process with the same intensity as in Scenario~I. The local estimators recover the true active set almost perfectly once $\mu$ is moderate: the average TPR of LSI increases from 87.5\% at $\mu=100$ to 100\% for $\mu\ge200$, and that of LLI from 63\% at $\mu=100$ to 92.5\% at $\mu=200$, 99\% at $\mu=300$ and 100\% for $\mu\ge500$. In contrast, the global estimators never exceed a TPR of 50.0\%. GLI increases from 26.5\% to 50.0\%, reaching exactly 50.0\% with zero standard deviation at $\mu=600$, which means that it selects exactly one of the two relevant predictors in every replication; GSI increases from 28.0\% to 47.5\%, and GALI attains only between 15.5\% and 31.0\%. This is expected, because each relevant predictor acts in only one quadrant, and a single constant coefficient cannot represent such an effect. The FPR is small for all methods. It is at most 0.6\% for LSI, and equal to 0.000 at $\mu=500$, and at most 1.4\% for LLI, while GALI has the largest FPR at every $\mu$ (3\%--5.1\%). At the event locations, the average local TPR is 91.4\% for LSI and 77.1\% for LLI at $\mu=100$; it equals 100.0\% for $\mu\ge200$ for LSI and for $\mu\ge500$ for LLI. The local FPR is at most 2.3\% for LLI and 1.7\% for LSI, and that of LSI decreases to 0.2\% at $\mu=600$. LSI also gives the most accurate coefficient surfaces at every $\mu$, with RMSPE$(\bm\beta)$ decreasing from 0.139 to 0.030, followed by LLI (from 0.214 to 0.127). The RMSPE of the global estimators remains at about 0.24--0.26 for every $\mu$. Compared with Scenario~I, the absence of clustering makes the problem easier for the local estimators: full recovery is reached at smaller $\mu$, the FPR is lower, and the RMSPE of LSI at $\mu=600$ is about half its Scenario~I value (0.030 against 0.059). As in the Scenario~I, LLI selects a larger $\lambda$ than GLI at every $\mu$, between two and four times as large, and for both estimators the selected $\lambda$ decreases overall with $\mu$ (from 0.0199 to 0.0123 for LLI and from 0.0084 to 0.0030 for GLI). For LSI the selected $\lambda$ is nearly constant (0.0199--0.0232), and for GSI it varies between 0.0129 and 0.0210 without a monotone trend. All methods run in under 0.25 seconds per fit on average, with LSI the slowest (0.13--0.24 seconds) and the remaining methods requiring at most 0.04 seconds. 

\begin{table}[!htbp]
\centering
\caption{Scenario~III (strong clustering, $\kappa=25$, $\sigma=0.0283$): mean (standard deviation) over $1000$ replications of the average of TPR, FPR, average local TPR and FPR, RMSPE of the coefficient surfaces and run time.}
\label{tab:sim-s3}
\resizebox{\linewidth}{!}{%
\begin{tabular}{llrrrrrrr}
\toprule
$\mu$ & Method & TPR & FPR & RMSPE$(\bm \beta)$ & $\overline{\operatorname{TPR}}_{\mathrm{loc}}$ & $\overline{\operatorname{FPR}}_{\mathrm{loc}}$ & $\lambda$ & Runtime (s)\\
\midrule
\multirow{5}{*}{100}
& LLI  & 0.633 (0.266) & 0.000 (0.001) & 0.075 (0.058) & 0.261 (0.182) & 0.004 (0.003) & 0.0091 (0.0079) & 0.80 (0.08) \\
& LSI  & 0.049 (0.067) & 0.063 (0.010) & 7.498 (2.370) & 0.018 (0.049) & 0.018 (0.003) & 0.0055 (0.0017) & 3.21 (0.70) \\
& GLI  & 0.571 (0.245) & 0.015 (0.021) & 0.336 (0.255) & --            & --            & 0.0015 (0.0011) & 0.39 (0.05) \\
& GSI  & 0.140 (0.113) & 0.072 (0.014) & 11.692 (4.555) & --           & --            & 0.0052 (0.0015) & 2.19 (0.49) \\
& GALI & 0.517 (0.224) & 0.016 (0.022) & 0.342 (0.235) & --            & --            & --              & 0.57 (0.06) \\
\midrule
\multirow{5}{*}{200}
& LLI  & 0.851 (0.197) & 0.002 (0.004) & 0.136 (0.077) & 0.458 (0.227) & 0.008 (0.004) & 0.0039 (0.0044) & 1.18 (0.43) \\
& LSI  & 0.124 (0.097) & 0.082 (0.016) & 5.646 (2.123) & 0.054 (0.075) & 0.023 (0.005) & 0.0062 (0.0023) & 4.70 (1.65) \\
& GLI  & 0.809 (0.152) & 0.060 (0.034) & 0.669 (0.290) & --            & --            & 0.0009 (0.0006) & 0.61 (0.24) \\
& GSI  & 0.166 (0.113) & 0.088 (0.017) & 6.812 (3.177) & --            & --            & 0.0069 (0.0042) & 4.05 (1.59) \\
& GALI & 0.819 (0.154) & 0.062 (0.032) & 0.639 (0.292) & --            & --            & --              & 0.87 (0.33) \\
\midrule
\multirow{5}{*}{300}
& LLI  & 0.958 (0.085) & 0.004 (0.005) & 0.190 (0.076) & 0.611 (0.196) & 0.011 (0.003) & 0.0018 (0.0011) & 1.78 (0.54) \\
& LSI  & 0.140 (0.101) & 0.092 (0.016) & 4.513 (1.329) & 0.054 (0.082) & 0.026 (0.004) & 0.0055 (0.0017) & 6.49 (1.87) \\
& GLI  & 0.807 (0.163) & 0.087 (0.043) & 0.773 (0.263) & --            & --            & 0.0007 (0.0006) & 0.92 (0.34) \\
& GSI  & 0.193 (0.118) & 0.096 (0.018) & 5.067 (1.720) & --            & --            & 0.0059 (0.0019) & 6.71 (2.49) \\
& GALI & 0.819 (0.143) & 0.088 (0.044) & 0.763 (0.266) & --            & --            & --              & 1.30 (0.46) \\
\midrule
\multirow{5}{*}{500}
& LLI  & 0.975 (0.048) & 0.006 (0.007) & 0.192 (0.059) & 0.706 (0.158) & 0.012 (0.003) & 0.0014 (0.0005) & 2.89 (0.51) \\
& LSI  & 0.174 (0.108) & 0.099 (0.017) & 3.740 (1.003) & 0.070 (0.096) & 0.028 (0.004) & 0.0047 (0.0013) & 9.73 (2.58) \\
& GLI  & 0.870 (0.104) & 0.156 (0.044) & 1.119 (0.289) & --            & --            & 0.0003 (0.0001) & 1.51 (0.30) \\
& GSI  & 0.244 (0.127) & 0.108 (0.018) & 4.831 (1.385) & --            & --            & 0.0045 (0.0011) & 11.57 (2.63) \\
& GALI & 0.873 (0.103) & 0.154 (0.043) & 1.129 (0.278) & --            & --            & --              & 2.17 (0.40) \\
\midrule
\multirow{5}{*}{600}
& LLI  & 0.983 (0.064) & 0.008 (0.007) & 0.220 (0.074) & 0.694 (0.183) & 0.013 (0.004) & 0.0013 (0.0006) & 3.11 (0.81) \\
& LSI  & 0.173 (0.113) & 0.097 (0.016) & 3.559 (0.924) & 0.052 (0.076) & 0.027 (0.004) & 0.0050 (0.0013) & 10.24 (4.07) \\
& GLI  & 0.862 (0.101) & 0.168 (0.042) & 1.205 (0.278) & --            & --            & 0.0002 (0.0001) & 1.61 (0.25) \\
& GSI  & 0.258 (0.120) & 0.108 (0.017) & 4.831 (1.430) & --            & --            & 0.0046 (0.0010) & 12.27 (2.54) \\
& GALI & 0.873 (0.111) & 0.173 (0.043) & 1.206 (0.268) & --            & --            & --              & 2.29 (0.33) \\
\bottomrule
\end{tabular}}
\end{table}
Table~\ref{tab:sim-s3} reports the results for Scenario~III. LLI attains the highest TPR at every $\mu$, increasing from 63.3\% at $\mu=100$ to 98.3\% at $\mu=600$. The TPR of GLI also increases broadly with $\mu$, from 57.1\% to between 81\% and 87\% for $\mu\ge200$, whereas both SCAD-type estimators recover few of the active predictors, with TPR below 26\% for GSI and below 18\% for LSI. LLI is also the most conservative method: its FPR does not exceed 0.8\% for any $\mu$. In contrast, the FPR of GLI grows from 1.5\% at $\mu=100$ to 16.8\% at $\mu=600$, and the SCAD-type estimators select between 6\% and 11\% of the inactive predictors throughout. GALI performs similarly to GLI: its TPR is lower at $\mu=100$ (51.7\% against 57.1\%) and slightly higher for $\mu\ge200$ (81.9--87.3\%), its FPR is nearly the same (1.6--17.3\%), and its RMSPE$(\bm\beta)$ differs from that of GLI by at most 0.03, so the adaptive weights bring little improvement in this scenario. LLI gives the most accurate coefficient surfaces, with the lowest RMSPE$(\bm\beta)$ for every $\mu$ (0.075--0.220), followed by GLI and GALI (0.336--1.205 and 0.342--1.206, respectively). The SCAD-type estimators have much larger RMSPE, which decreases with $\mu$ from 7.50 to 3.56 for LSI and from 11.69 to 4.83 for GSI. At the local level, the average local TPR of LLI increases from 26.1\% at $\mu=100$ to about 70\% for $\mu\ge500$, with local FPR at most 1.3\%, whereas LSI recovers at most 7.0\% of the locally active predictors. Under strong clustering, recovering the location-specific active set is therefore considerably harder than recovering the global one, even for LLI. For the LASSO-type estimators the selected $\lambda$ decreases as $\mu$ increases: from 0.0015 to 0.0002 for GLI and from 0.0091 to 0.0013 for LLI. This is expected, since the log-likelihood in CBIC grows with the number of points while the complexity term grows only logarithmically, so less shrinkage is needed when more points are available; the smaller $\lambda$ of GLI is also consistent with its increasing FPR. The local estimator selects a larger $\lambda$ than its global counterpart (for example, 0.0091 against 0.0015 at $\mu=100$), because the localized design contains $R$ columns for each predictor and a stronger penalty is required to keep the model sparse. For the SCAD-type estimators $\lambda$ varies little with $\mu$, lying between 0.0045 and 0.0069, and LSI and GSI select penalties of similar magnitude. For GALI, $\lambda$ is not reported because its penalty is weighted by the adaptive weights, so its value is not on the same scale as for the other methods. LLI requires 0.8--3.1 seconds per fit, roughly twice the time of GLI (0.4--1.6 seconds), owing to the larger localized design, while GALI requires 0.6--2.3 seconds. The SCAD-type estimators are the most expensive, requiring 3.2--10.2 seconds for LSI and 2.2--12.3 seconds for GSI. Overall, in this scenario LLI is the best method for active-variable detection at both the global and the local level and for coefficient-surface recovery, while the SCAD-type estimators perform poorly. Across the three scenarios, LLI consistently recovers the locally active predictors and the coefficient surfaces better than the global methods, and LSI does so in Scenarios~I and~II; for both local estimators the selection accuracy increases with $\mu$. The poor performance of the SCAD-type estimators in Scenario~III stems from the combination of a rank-deficient design ($p=500$ predictors on $256$ pixels), strong clustering, and low-variance irrelevant predictors. Because the SCAD penalty is constant for $|\beta|>\breve{\tau}\cdot \lambda$ where the SCAD parameter, $\breve{\tau} > 2$, an irrelevant predictor that enters the model can attain an arbitrarily large coefficient at no additional cost and is then used to fit the clusters, whereas the LASSO penalty grows linearly in $|\beta|$ and prevents this. When $p$ is small relative to the number of pixels, as in Scenarios~I and~II, the non-convex problem is well behaved and LSI attains the best performance.
\section{Analysis of St. Louis Data}\label{s:real}
In this section we describe the data which motivates us to propose this novel methodology of local variable selection. 
\begin{figure}[h!]
    \centering
    \includegraphics[width=0.5\linewidth]{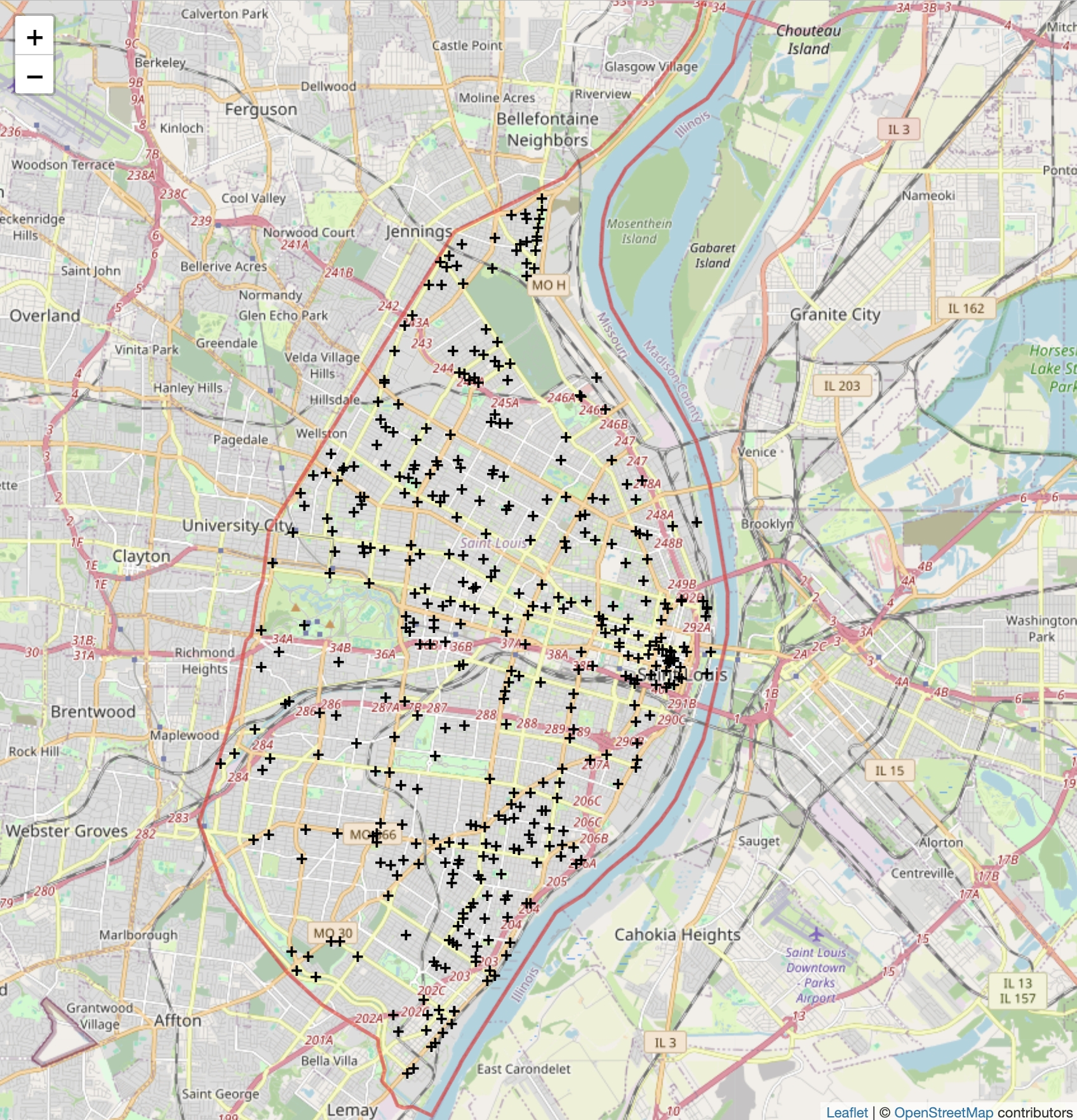}
    \caption{Incidence points of aggravated assault crimes in St. Louis City are marked by ``+''.}
    \label{fig:obs_data}
\end{figure}
\begin{figure}[h!]
    \centering
    \includegraphics[width=0.6\linewidth]{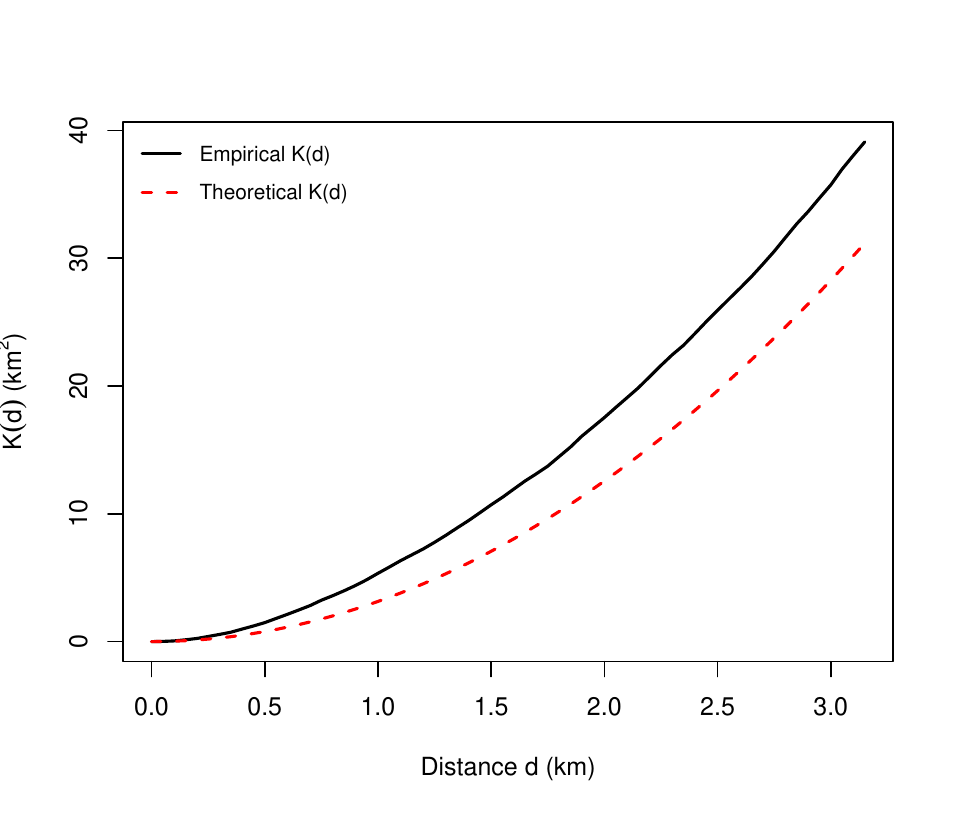}
    \caption{Empirical and theoretical $K-$function of crime occurrences in St. Louis. The black solid line is corresponding to the Empirical $K-$function and the red dotted line denotes the theoretical $K-$function.}
    \label{fig:k_fun}
\end{figure}
In Figure~\ref{fig:obs_data} we visualize the spatial distribution of violent crime occurrences in the city of St. Louis. This data set considers the incidence of assaults, which consists of assaults in different degrees, murders, harassments, and so on. Mainly, this type of assault incident is considered as an aggravated assault. The data is accessible at the following link \url{https://www.crimemapping.com/Share/66dfeeb3d9874dc78fdf8be1ad7f0c4c}. From the first of March to the $30$th April in the year, $2026$, St. Louis city has experienced $652$ aggravated assault incidents, which is quite a serious concern for St. Louis city and the data is visualized in Figure~\ref{fig:obs_data}. After removing the duplicate locations there exists $405$ unique incidence points during two months. Before diving into the detailed variable selection results we first discuss the second order properties of the observed incidence points. Figure~\ref{fig:k_fun} illustrates the difference between empirical Ripley’s K-function and theoretical $K-$function assuming that the points are the realizations of a homogeneous Poisson point process and its theoretical $K-$function is $K(d) = \pi d^2$. In Figure~\ref{fig:k_fun}, we detect that the empirical K-function consistently lies above the theoretical $K-$function curve along the entire range of spatial distances, $d$. In this way Figure~\ref{fig:k_fun} indicates a departure from a homogeneous Poisson point process, and such departure may arise from spatial variation in the first-order intensity, second-order dependence, or both. The departure in upward direction exhibits that the observed incidence points have more neighboring incidence points within distance $d$ than that would be expected under the homogeneous Poisson process assumption, and this difference increases with $d$. This departure from the assumption of a homogeneous intensity function motivates us to incorporate spatial dynamics into the intensity function model. For successful inclusion of spatial dynamism, we require selecting variables locally when there are many predictors; otherwise, the inherited bias in the predictors will increase bias in intensity estimation.

The choice of a feature set for modeling the inhomogeneous intensity function is based solely on established principles of environmental criminology and crime pattern theory, rather than on subjective preference (\cite{brantingham2017environment}). Specifically, we consider cardinality of the following predictor variables, such as, nearby road types (motorways, primary, secondary, tertiary, residential, and service roads), urban morphology (building density), land use (commercial, industrial, and retail areas), entertainment and nightlife activities (bars, pubs, nightclubs, and cinemas), food and hospitality establishments (restaurants, fast-food outlets, and cafés), recreational facilities (parks and playgrounds), public transportation infrastructure (bus stops), vehicle-related infrastructure (parking facilities and fuel stations), retail amenities (convenience stores and supermarkets), financial services (banks and ATMs), and healthcare facilities (hospitals and pharmacies) within the multiple spatial scales, 50, 100, 200, 500, and 1000 meters around each event location. These features generally provide an idea of the surroundings of crime spots, which influence accessibility, routine human activities, population concentration, land-use intensity, and opportunities for offender–victim interactions. These features are conventionally recognized as key factors of the spatial distribution of violent crimes. This multiscale spatial representation of the feature space makes the model effectively accountable for both immediate neighborhood effects and broader environmental influences on the intensity of aggravated assault.

\begin{table*}[!htbp]
\centering
\caption{Comparison of selected variables. A check mark ($\checkmark$) indicates that the corresponding predictor was selected by the method, whereas ($\times$) indicates that it was not selected.}
\label{tab:variable_selection}
\renewcommand{\arraystretch}{1.00}
\setlength{\tabcolsep}{3pt}
\resizebox{0.6\textwidth}{!}{%
\begin{tabular}{llccccc}
\toprule
\textbf{Category} & \textbf{Predictor} & \textbf{GLI} & \textbf{GSI} & \textbf{GALI} & \textbf{LLI} & \textbf{LSI} \\
\midrule
\multirow{12}{*}{Road Types}
& Intersection nodes within 50m  & $\times$ & $\times$ & $\times$ & $\times$ & $\checkmark$ \\
& Motorway within 50m            & $\checkmark$ & $\times$ & $\checkmark$ & $\times$ & $\times$ \\
& Motorway within 500m           & $\times$ & $\times$ & $\times$ & $\times$ & $\checkmark$ \\
& Primary road within 50m        & $\checkmark$ & $\times$ & $\times$ & $\times$ & $\times$ \\
& Primary road within 200m       & $\checkmark$ & $\checkmark$ & $\checkmark$ & $\times$ & $\checkmark$ \\
& Primary road within 500m       & $\checkmark$ & $\checkmark$ & $\checkmark$ & $\times$ & $\checkmark$ \\
& Primary road within 1000m      & $\times$ & $\times$ & $\times$ & $\times$ & $\checkmark$ \\
& Residential road within 50m    & $\checkmark$ & $\checkmark$ & $\checkmark$ & $\times$ & $\checkmark$ \\
& Residential road within 500m   & $\times$ & $\times$ & $\times$ & $\checkmark$ & $\checkmark$ \\
& Residential road within 1000m  & $\checkmark$ & $\checkmark$ & $\checkmark$ & $\times$ & $\times$ \\
& Secondary road within 50m      & $\checkmark$ & $\checkmark$ & $\checkmark$ & $\times$ & $\times$ \\
& Secondary road within 500m     & $\checkmark$ & $\checkmark$ & $\checkmark$ & $\times$ & $\times$ \\
\midrule
\multirow{32}{*}{Buildings / Land Use}
& Buildings within 50m           & $\checkmark$ & $\checkmark$ & $\times$ & $\times$ & $\times$ \\
& Buildings within 100m          & $\times$ & $\checkmark$ & $\times$ & $\checkmark$ & $\checkmark$ \\
& Buildings within 1000m         & $\times$ & $\times$ & $\times$ & $\checkmark$ & $\checkmark$ \\
& Commercial within 100m         & $\checkmark$ & $\checkmark$ & $\times$ & $\times$ & $\times$ \\
& Industrial within 50m          & $\times$ & $\times$ & $\times$ & $\times$ & $\checkmark$ \\
& Industrial within 500m         & $\checkmark$ & $\checkmark$ & $\checkmark$ & $\times$ & $\times$ \\
& Retail within 50m              & $\checkmark$ & $\checkmark$ & $\checkmark$ & $\times$ & $\checkmark$ \\
& Retail within 200m             & $\checkmark$ & $\times$ & $\checkmark$ & $\times$ & $\checkmark$ \\
& Retail within 500m             & $\checkmark$ & $\checkmark$ & $\checkmark$ & $\times$ & $\times$ \\
& Retail within 1000m            & $\times$ & $\times$ & $\times$ & $\times$ & $\checkmark$ \\
& Bars within 50m                & $\times$ & $\times$ & $\times$ & $\checkmark$ & $\checkmark$ \\
& Pubs within 500m               & $\checkmark$ & $\checkmark$ & $\checkmark$ & $\times$ & $\checkmark$ \\
& Pubs within 1000m              & $\times$ & $\times$ & $\times$ & $\times$ & $\checkmark$ \\
& Nightclubs within 200m         & $\times$ & $\times$ & $\times$ & $\times$ & $\checkmark$ \\
& Nightclubs within 1000m        & $\checkmark$ & $\checkmark$ & $\checkmark$ & $\times$ & $\times$ \\
& Restaurants within 50m         & $\checkmark$ & $\checkmark$ & $\checkmark$ & $\times$ & $\times$ \\
& Fast food within 50m           & $\times$ & $\times$ & $\times$ & $\times$ & $\checkmark$ \\
& Fast food within 100m          & $\times$ & $\times$ & $\times$ & $\times$ & $\checkmark$ \\
& Fast food within 200m          & $\checkmark$ & $\checkmark$ & $\times$ & $\times$ & $\times$ \\
& Fast food within 1000m         & $\checkmark$ & $\checkmark$ & $\checkmark$ & $\times$ & $\checkmark$ \\
& Cafes within 100m              & $\checkmark$ & $\checkmark$ & $\checkmark$ & $\checkmark$ & $\checkmark$ \\
& Cafes within 200m              & $\times$ & $\times$ & $\times$ & $\times$ & $\checkmark$ \\
& Cinema within 500m             & $\times$ & $\times$ & $\times$ & $\checkmark$ & $\checkmark$ \\
& Convenience within 200m        & $\checkmark$ & $\checkmark$ & $\checkmark$ & $\checkmark$ & $\checkmark$ \\
& Convenience within 500m        & $\checkmark$ & $\checkmark$ & $\checkmark$ & $\times$ & $\times$ \\
& Convenience within 1000m       & $\times$ & $\times$ & $\times$ & $\checkmark$ & $\checkmark$ \\
& Supermarket within 200m        & $\checkmark$ & $\times$ & $\times$ & $\times$ & $\checkmark$ \\
& Supermarket within 500m        & $\times$ & $\times$ & $\times$ & $\times$ & $\checkmark$ \\
& Fuel station within 100m       & $\times$ & $\times$ & $\times$ & $\checkmark$ & $\times$ \\
& Fuel station within 200m       & $\times$ & $\times$ & $\times$ & $\times$ & $\checkmark$ \\
& Fuel station within 500m       & $\checkmark$ & $\checkmark$ & $\checkmark$ & $\times$ & $\times$ \\
& Fuel station within 1000m      & $\checkmark$ & $\checkmark$ & $\checkmark$ & $\times$ & $\checkmark$ \\
\midrule
\multirow{3}{*}{Transportation}
& Bus stops within 1000m         & $\times$ & $\times$ & $\times$ & $\times$ & $\checkmark$ \\
& Parking within 50m             & $\checkmark$ & $\times$ & $\checkmark$ & $\times$ & $\times$ \\
& Parking within 1000m           & $\times$ & $\times$ & $\times$ & $\checkmark$ & $\times$ \\
\midrule
\multirow{5}{*}{Education}
& Schools within 500m            & $\times$ & $\times$ & $\times$ & $\checkmark$ & $\checkmark$ \\
& Schools within 1000m           & $\times$ & $\checkmark$ & $\times$ & $\times$ & $\checkmark$ \\
& Universities within 50m        & $\checkmark$ & $\times$ & $\checkmark$ & $\times$ & $\checkmark$ \\
& Universities within 100m       & $\checkmark$ & $\checkmark$ & $\checkmark$ & $\times$ & $\times$ \\
& Universities within 500m       & $\times$ & $\times$ & $\times$ & $\checkmark$ & $\checkmark$ \\
\midrule
\multirow{5}{*}{Healthcare}
& Hospital within 50m            & $\checkmark$ & $\checkmark$ & $\checkmark$ & $\times$ & $\times$ \\
& Hospital within 100m           & $\checkmark$ & $\times$ & $\times$ & $\times$ & $\times$ \\
& Hospital within 200m           & $\times$ & $\times$ & $\times$ & $\times$ & $\checkmark$ \\
& Hospital within 1000m          & $\times$ & $\times$ & $\times$ & $\checkmark$ & $\times$ \\
& Pharmacy within 500m           & $\times$ & $\times$ & $\times$ & $\times$ & $\checkmark$ \\
\midrule
\multirow{6}{*}{Recreation}
& Parks within 50m               & $\times$ & $\times$ & $\times$ & $\checkmark$ & $\checkmark$ \\
& Parks within 100m              & $\checkmark$ & $\checkmark$ & $\checkmark$ & $\times$ & $\times$ \\
& Parks within 200m              & $\checkmark$ & $\checkmark$ & $\checkmark$ & $\times$ & $\times$ \\
& Playgrounds within 50m         & $\times$ & $\times$ & $\times$ & $\times$ & $\checkmark$ \\
& Playgrounds within 100m        & $\checkmark$ & $\times$ & $\times$ & $\checkmark$ & $\checkmark$ \\
& Playgrounds within 500m        & $\times$ & $\times$ & $\times$ & $\times$ & $\checkmark$ \\
\bottomrule
\end{tabular}%
}
\end{table*}
In this real data analysis, we fix $J=3$. Table~\ref{tab:variable_selection} shows a substantial difference between the variables selected by the global and the localized methods. LSI retains the largest number of predictors ($40$), whereas LLI yields the sparsest model with only $15$ predictors. The global methods GLI, GSI and GALI select $33$, $27$ and $26$ predictors, respectively. GLI and LSI also retain the largest number of micro-scale features, measured within the immediate spatial buffers of 50--200\,m ($22$ and $21$, respectively), followed by GSI and GALI ($15$ each) and LLI ($7$). Only two predictors, cafes within 100\,m and convenience stores within 200\,m, are selected by all five methods. The main difference between the global and localized methods lies in the spatial scales at which environmental features are retained. For example, the global methods select convenience stores within 200\,m and 500\,m, whereas both localized methods select them within 200\,m and 1000\,m. Similarly, GLI, GSI and GALI select nightclubs within 1000\,m, whereas LSI selects nightclubs within 200\,m. The scale-specific differences are also evident for schools: GSI selects this predictor within 1000\,m radius, whereas LLI selects the predictor at 500\,m radius, and LSI selects both the radius 500\,m and 1000\,m for this predictor. For hospitals, the global methods retain only short-range predictors. The global models, GLI, GSI and GALI select hospitals within 50\,m scale and GLI additionally selects hospitals at the 100\,m scale, whereas LSI selects the hospitals at 200\,m scale and LLI selects that predictor at 1000\,m scale. For universities, all global methods select the 100\,m scale, while LLI selects the predictor at the 500\,m scale and LSI selects the predictor at both the 50\,m and 500\,m scales. Several predictors are selected only by the localized methods, including bars and parks within 50\,m, buildings within 1000\,m, cinemas within 500\,m and residential roads within 500\,m, all of which are retained by both LLI and LSI. LSI additionally retains a broad collection of short-range (50--200\,m) predictors, including intersections, residential roads, industrial and retail areas, nightclubs, fast-food outlets, cafes, supermarkets, fuel stations, hospitals and playgrounds. LLI produces a relatively sparse predictor set, consisting of residential roads, buildings, bars, cafes, cinemas, convenience stores, fuel stations, parking, schools, universities, hospitals, parks and playgrounds, and $12$ of its $15$ predictors are also selected by LSI. Short-range selection is not unique to the localized procedures, however, since GLI also retains a substantial number of micro-scale predictors. The location-specific selection maps show whether these localized combinations of features and scales vary across different parts of St.~Louis.
\begin{figure}[!htbp]
    \centering
    \includegraphics[
        width=0.9\linewidth, trim=0 50 0 0,
        clip]{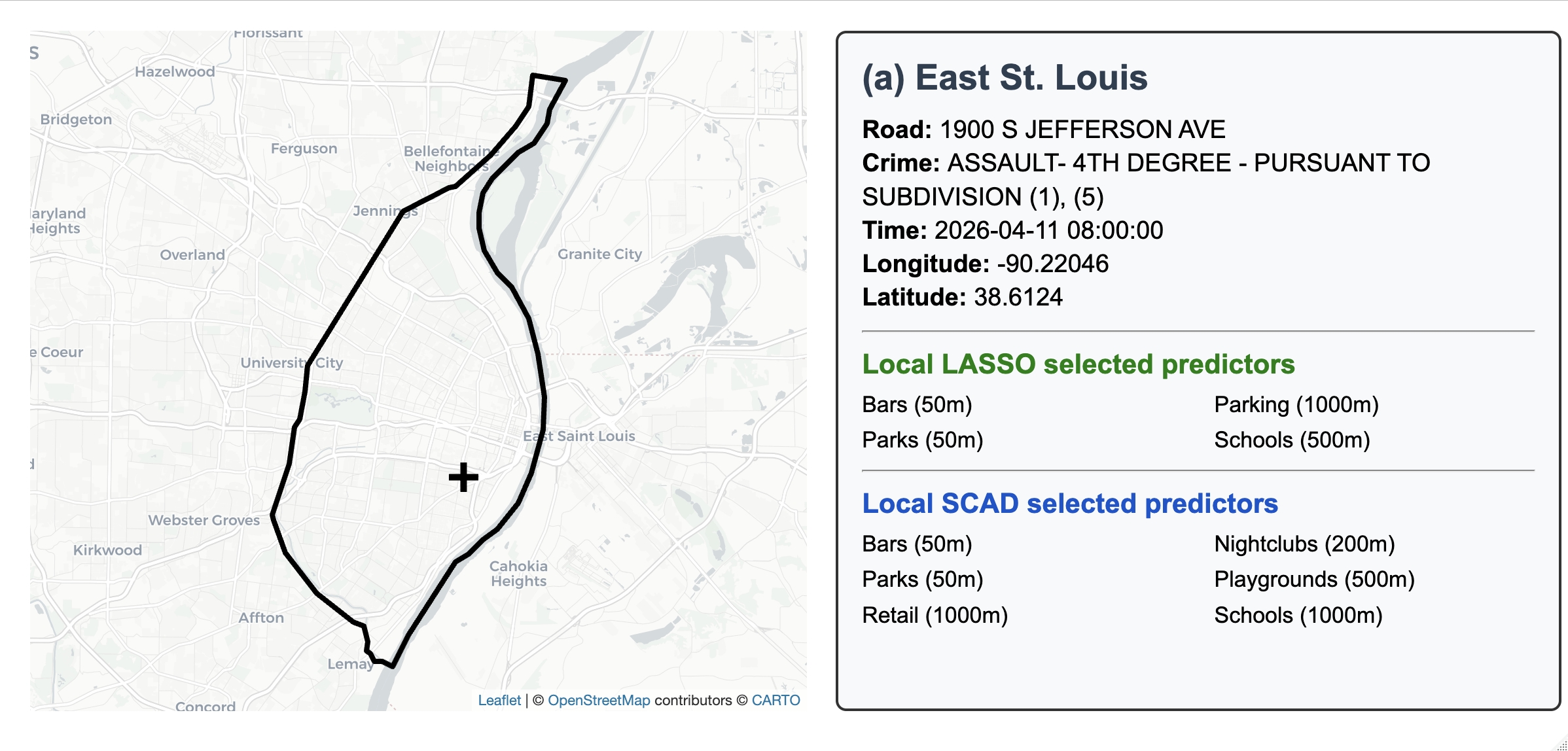}
    \caption{Selected predictor set in the eastern part of St.~Louis by LLI and LSI.}
    \label{fig:east_stl}
\end{figure}
\begin{figure}[!htbp]
    \centering
\includegraphics[
        width=0.9\linewidth,
        trim=0 50 0 0,
        clip
    ]{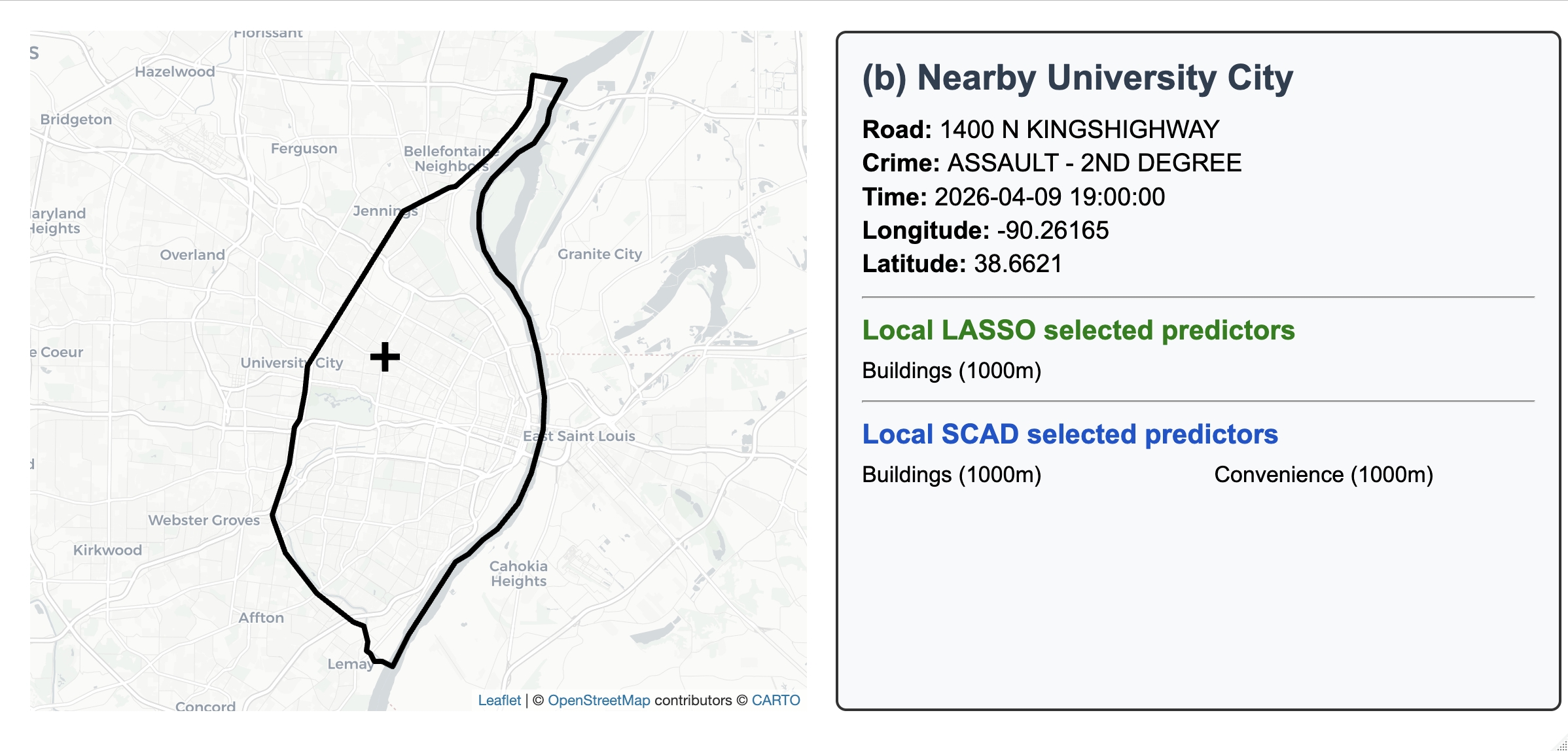}
    \caption{Selected Predictor set near University City of St. Louis by LLI and LSI.}
    \label{fig:univ_stl}
\end{figure}
\begin{figure}[!htbp]
    \centering
\includegraphics[
        width=0.9\linewidth,
        trim=0 50 0 0,
        clip
    ]{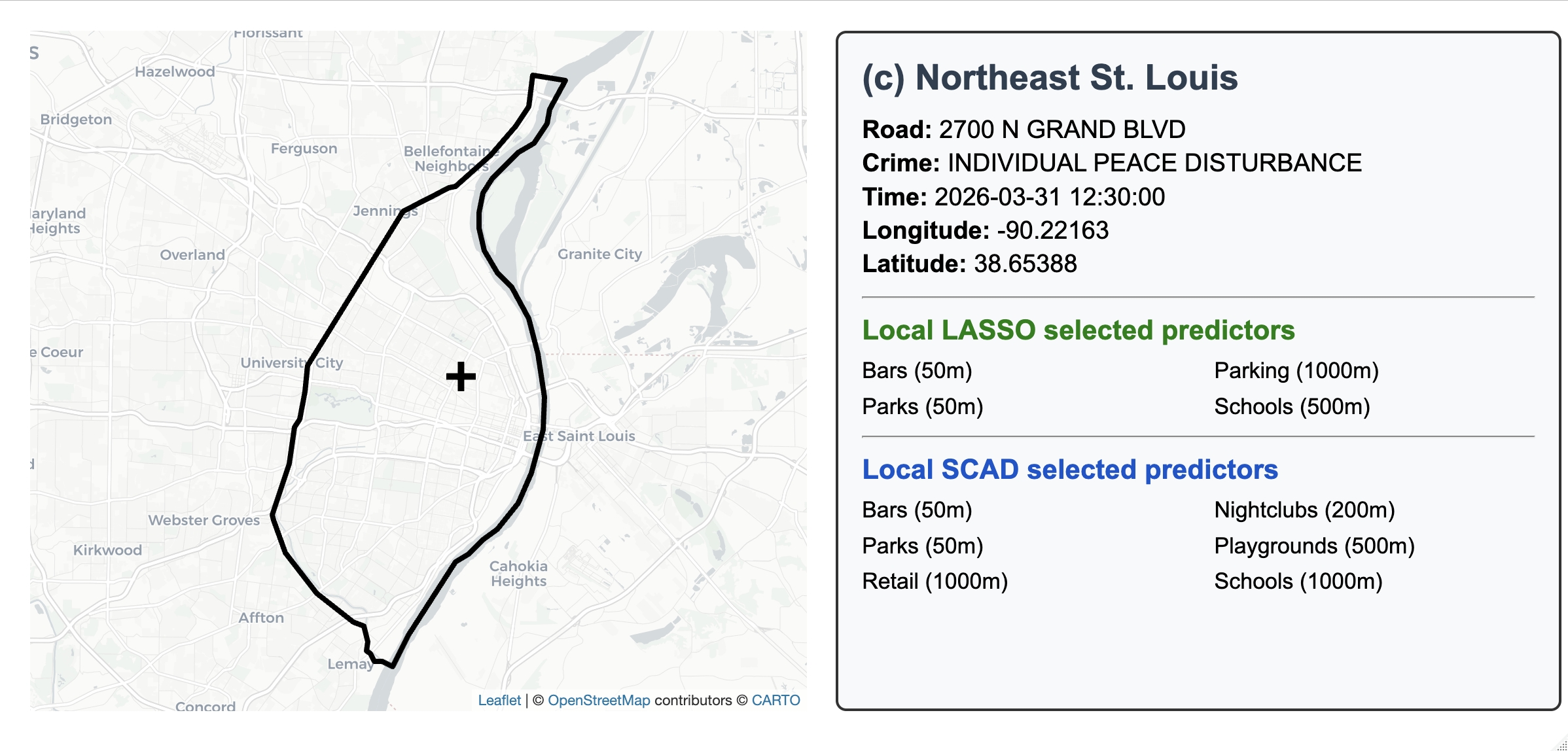}
    \caption{Selected Predictor set in North-Eastern part of St. Louis by LLI and LSI.}
    \label{fig:north_east_stl}
\end{figure}
\begin{figure}[!htbp]
    \centering
\includegraphics[
        width=0.9\linewidth,
        trim=0 50 0 0,
        clip
    ]{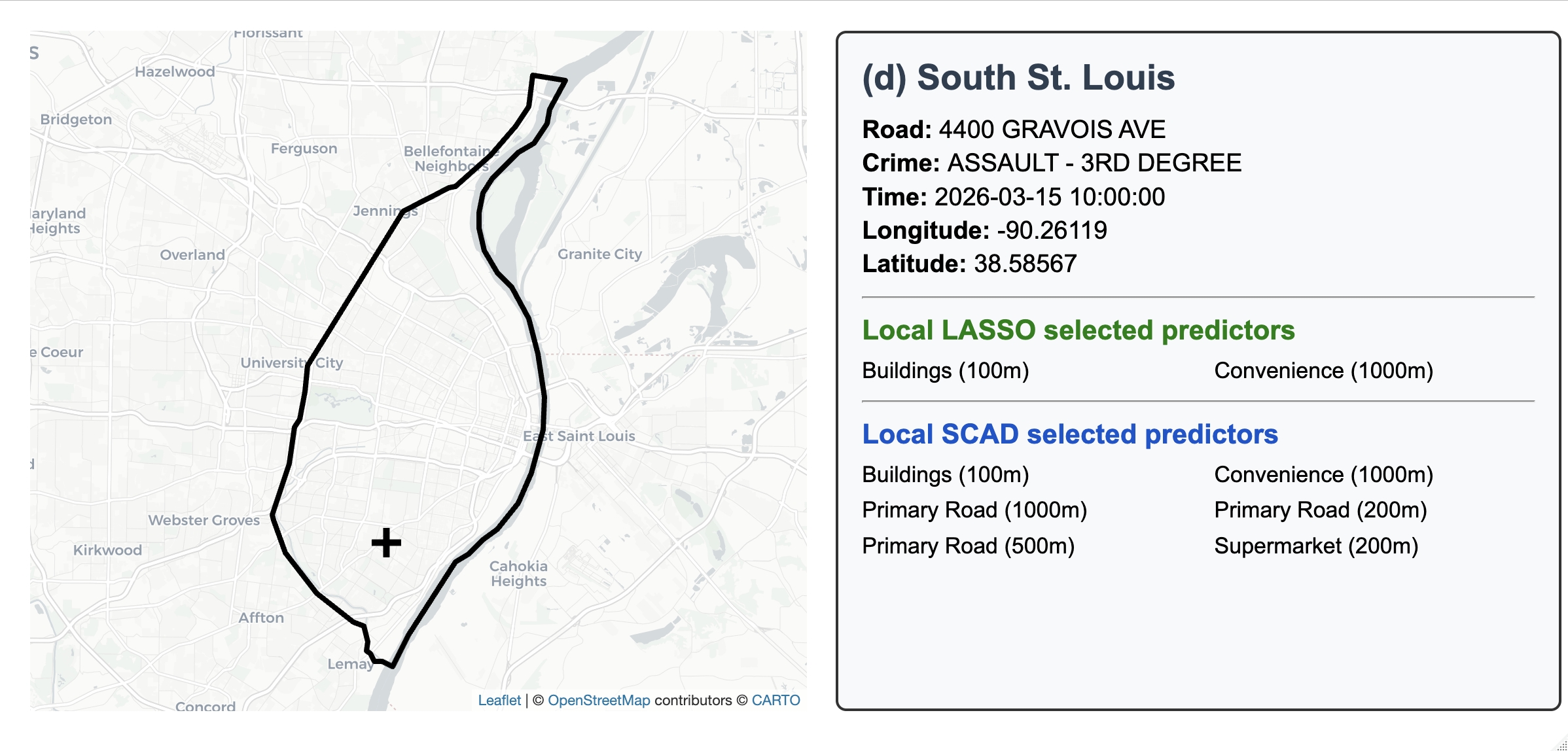}
    \caption{Selected Predictor set in Southern part of St. Louis by LLI and LSI.}
    \label{fig:south_stl}
\end{figure}
In Figures~\ref{fig:east_stl}--\ref{fig:south_stl}, we further illustrate how the spatial heterogeneity in the estimated activation set, captured by the proposed localized variable-selection methods at three representative locations in St.~Louis. In the eastern and northeastern parts of the study region, LLI selects a relatively sparse set of predictors, namely bars within 50\,m, parks
within 50\,m, 1000\,m and schools within 500\,m. LSI identifies a richer local structure at these locations, retaining bars and parks within 50\,m and schools within 1000\,m, while additionally selecting nightclubs within 200\,m, and playgrounds within 500\,m. However, local methods follow a distinctly different selection pattern in the
southern part of St.~Louis. At this location, LLI primarily identifies Buildings within 100\,m and convenience stores within 1000\,m, whereas LSI selects Buildings within 100\,m radius, convenience stores within 1000\,m radius, super market within 200\,m radius, and primary road within 200\,m, 500\,m, and 1000\,m radius. However from Figure~\ref{fig:univ_stl} we detect that nearby university city LLI selects only Buildings within 1000\,m radius whereas LSI selects Buildings and Convenience stores within 1000\,m radius. These location-specific differences in the selection set demonstrate how the selected predictor set varies across spatial scales. In particular, the richer predictor sets obtained by LSI, whereas the comparatively sparse predictor sets obtained by LLI emphasize a smaller set of locally dominant features.
\begin{figure}[!htbp]
    \centering
    \includegraphics[width=0.9\linewidth]{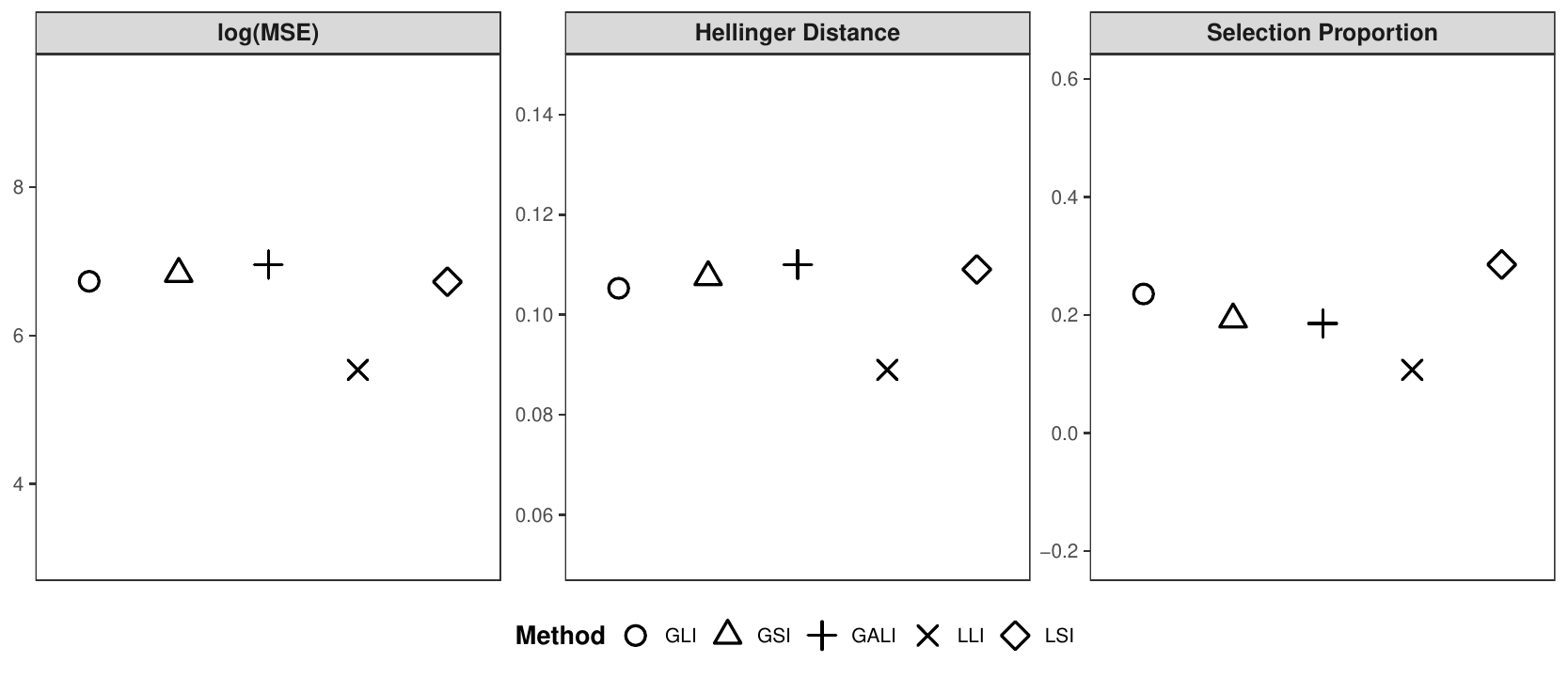}
    \caption{Comparison of predictive performance by comparing MSE and Hellinger Distance.}
    \label{fig:performance}
\end{figure}
To evaluate and compare the predictive accuracy of the competing models, we employ an un-penalized composite likelihood is used as post-selection refitting. Regularization methods (GLI, GSI, GALI, LLI, LSI) are effective for variable selection at the global and local levels, but in this section, we also compare their predictive performance. Once the predictors are selected at the global and local level, we use those predictors and standardize them to estimate the point-process intensity $\hat{\pi}^{(l)}(\bm u)$ for method $l$ using the unpenalized log-composite-likelihood of the Poisson process using the Berman-Turner quadrature scheme described in Section~\ref{ss:computation}.
\begin{figure}[!htbp]
\centering
\includegraphics[width=\linewidth]{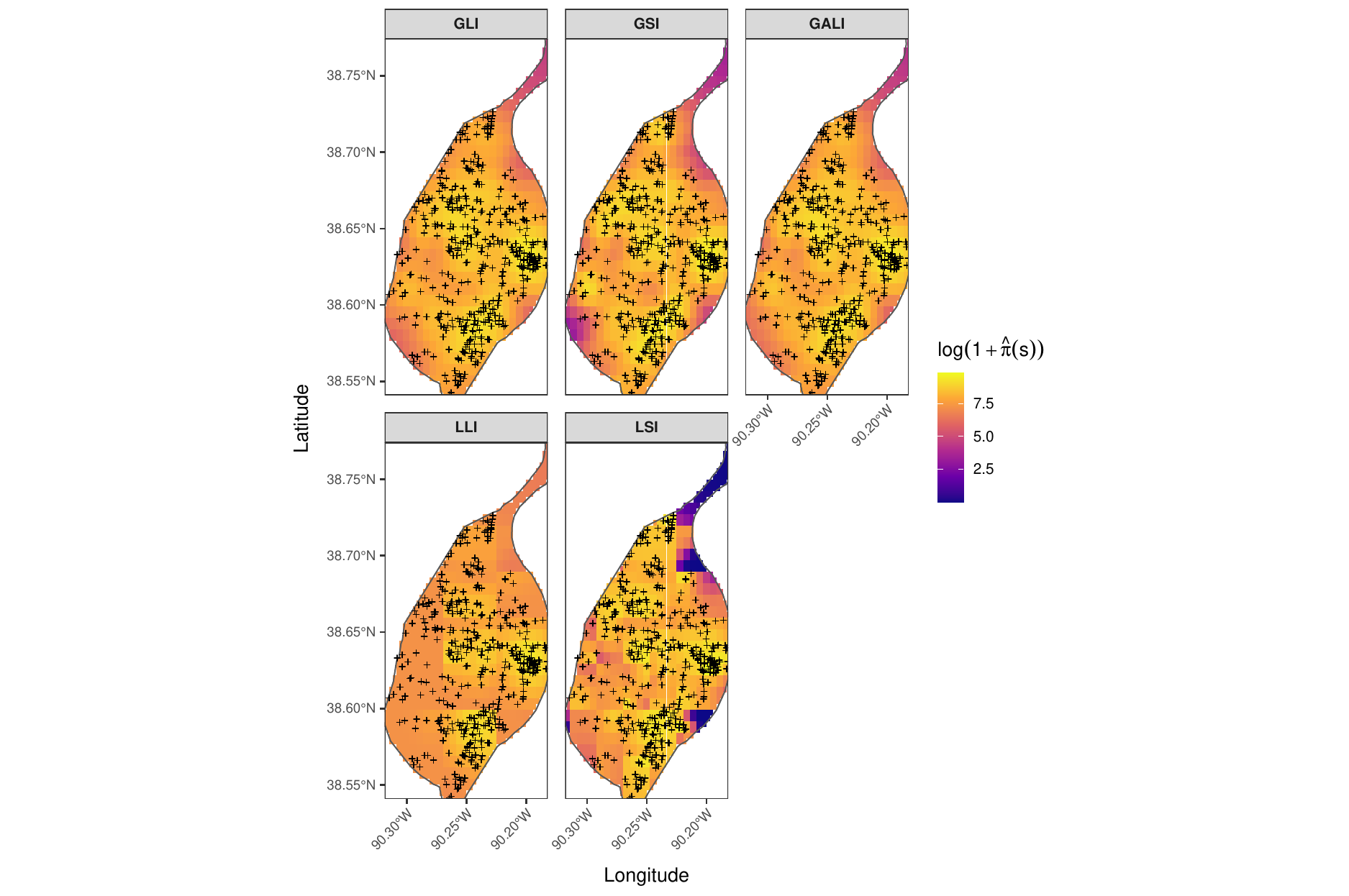}
\caption{Predicted intensity surface of all methods and ``+'' signs indicates the observed incidence points.}
\label{fig:est-intensity}
\end{figure}
To assess how well each method reproduces the observed pattern, we also compare observed and predicted counts. After selection, the Poisson model with only the selected columns is refitted without penalty, giving the fitted intensity $\widehat\pi_l(\cdot)$ of method $l$. Let $\bm q_m$, $m=1,\ldots,N$, denote the quadrature points introduced in Section~\ref{ss:computation}. Within each grid cell $\square_\xi$, we compute the observed point count  and the corresponding predicted expected count under method $l$ as
\[
N_\xi
=
\sum_{m:\bm q_m\in\square_\xi}\Delta_m,
\qquad
\widehat{\mu}_{\xi}^{(l)}
=
\sum_{m:\bm q_m\in\square_\xi}
\omega_m\widehat{\pi}_l(q_m),
\qquad \xi=1,\ldots,G,
\]
where $\omega_m$ and $\Delta_m$ denote the quadrature weight and event 
indicator associated with $\bm q_m$, respectively, as described in 
Section~\ref{ss:computation}. After that based on $N_\xi$ we prepare the set of all possible counts, $\mathscr{K}$ and from $l$th method we get predicted count, in each category based on $\widehat{\mu}_{\xi}^{(l)}$. Then the MSE corresponding to the $l$ th method becomes
\begin{equation}\label{eq:mse}
\text{MSE}^{(l)} = \frac{1}{[[\mathscr{K}]]} \sum_{k\in \mathscr{K}} \left(\left[\sum_{\xi=1}^G\mathbf{I}(N_\xi = k)\right] - \left[\sum_{\xi=1}^G\mathbf{I}(\hat{N}_{\xi}^{(l)} = k)\right]\right)^2,\;
\end{equation}
The discretized predicted counts in \eqref{eq:mse} is $\hat{N}_{\xi}^{(l)} = \lfloor\hat{\mu}_{\xi}^{(l)}\rceil$ where, $\lfloor v\rceil$ indicates the nearest integer to $v$, and observed counts $N_\xi$ are tabulated across common support categories $k \in \mathscr{K}$. The lower MSE indicates that $l$th method predicts the number of occurrences more correctly. 
Model performance is assessed using the MSE from \eqref{eq:mse}. We also calculate Category-wise Hellinger Distance ($H^{(l)}$) to compare the agreement between the empirical and model-specific grid-count distributions. From the probability vectors $\hat{\mathscr{P}}^{(l)} = (\hat{\tilde q}^{(l)}_{k})_{k \in \mathscr{K}}$ and $\mathscr{P} = (\tilde{q}_k)_{k \in \mathscr{K}}$ where $\mathscr{K}$ is the set of distinct counts i.e. $\{0,1,\dots,12\}$ observed in a grid and, $\tilde q_k = \frac{1}{G}\sum_{\xi=1}^{G} \mathbf{I}(N_\xi = k)$. The Hellinger distance is defined as:
    \begin{equation*}
        H^{(l)} = H\left(\mathscr{P}, \hat{\mathscr{P}}^{(l)}\right) = \frac{1}{\sqrt{2}} \sqrt{\sum_{k \in \mathscr{K}} \left(\sqrt{\tilde q_k} - \sqrt{\hat{\tilde{q}}^{(l)}_{k}}\right)^2},\; \mathrm{where}\; \hat{\tilde{q}}^{(l)}_{k} = \frac{1}{G}\sum_{\xi=1}^G \exp\{-\widehat{\mu}_{\xi}^{(l)}\}\frac{\{\widehat{\mu}_{\xi}^{(l)}\}^k}{k!}
    \end{equation*}
Since $0\leq  H\left(\mathscr{P}, \hat{\mathscr{P}}^{(l)}\right) \leq1$, where $ H\left(\mathscr{P}, \hat{\mathscr{P}}^{(l)}\right)=0$ corresponds to identical grid-count distributions, lower values of both $\operatorname{MSE}^{(l)}$ and $ H\left(\mathscr{P}, \hat{\mathscr{P}}^{(l)}\right)$ indicate better performance according to their respective measures. As shown in Figure~\ref{fig:performance}, LLI has the smallest MSE and the smallest Hellinger distance (HD). Along with the previous observation, LLI also shows higher sparsity w.r.t. LSI and also yields lowest MSE and HD values. Thus, Figure~\ref{fig:performance} shows that LLI is the most parsimonious model among the regularized methods. In Figure~\ref{fig:est-intensity}, we also compare the predicted intensity surface. This predicted intensity surface will provide insight into the spatial heterogeneity of the estimated intensity function. From Figure~\ref{fig:est-intensity}, we observe that almost all the methods recover the major spatial trend of assault incidents, with relatively high intensities concentrated in the central and southern parts of St. Louis city, whereas lower intensities are observed near the northern boundary. In particular, LSI identifies several localized regions of reduced intensity, providing evidence of local spatial heterogeneity in the covariate effects, which are less stressed under the global models. LLI provides the most parsimonious representation in the present application, retaining only 15 predictors. Although both LLI and LSI accommodate spatially varying covariate effects, LLI produces a substantially sparser selected set. LLI provides the smallest MSE and HD values in Figure~\ref{fig:performance}. Thus, LLI provides a favorable balance between sparsity and predictive accuracy in this application. Thus, the real data analysis provides empirical evidence that the associations between built-environment characteristics and aggravated-assault intensity are spatially heterogeneous across multiple spatial scales. The localized procedures reveal location-specific predictor sets that cannot be represented by a single global selection pattern. Among the localized procedures, LLI provides the most parsimonious representation of this heterogeneity.

\section{Conclusions}
In our research both the local methods, LLI and LSI consistently reveal that the set of relevant explanatory variables changes from one location to another. Although several predictors, including bars, building density, and public transportation variables, are selected across multiple locations, and the considerable variation is observed in the remaining set of predictors. Thus these results support the existence of heterogeneity in the spatial distribution of the feature set and demonstrate that local penalized regression can uncover neighborhood-specific associated predictors that would not be possible by global variable selection methods. In short, from the above discussions, we can see the localized models (LLI, LSI) preserve those broad effects while revealing \emph{where} these predictors actually matter for spatial incidences. In future research, we plan to extend the spatial variable selection perspective in the spatio-temporal direction, especially for the self-exciting process, and we will also consider different types of crimes incidences. The proposed methodology is developed within an inhomogeneous Poisson point process framework and primarily addresses spatial variation in the first-order intensity function. But via exploratory K-function analysis we observe the presence of substantial aggregation relative to complete spatial randomness, which may also reflect residual second-order dependence not explicitly represented by the current model. For the methodological advancements, we have a plan to extend this regularized variable selection method for a spatial clustered point process, for example, Cox process. We have a plan to establish the asymptotic properties of the regularized intensity estimator in the future, and also extend this local variable selection method for spatial point process in a linear network.
\section*{Data and Code Availability} The package is available at \url{https://github.com/debjoythakur/localvspp} and can be installed with \url{remotes::install_github("debjoythakur/localvspp")}.
\section*{Conflicts of interest}
The authors declare that they have no competing interests.
\section*{Author contributions statement}
The authors report there are no competing interests to declare. DT contribution: Data preprocessing, developing, and applying the methodology to St. Louis data. SNL contributes to the formulation of the modeling and methodological framework.
\section*{Acknowledgments}
The authors acknowledge support by NSF Grants: CMMI2235457.
\bibliographystyle{plainnat}
\bibliography{reference}

@article{zhang2010regularization,
  title={Regularization parameter selections via generalized information criterion},
  author={Zhang, Yiyun and Li, Runze and Tsai, Chih-Ling},
  journal={Journal of the American statistical Association},
  volume={105},
  number={489},
  pages={312--323},
  year={2010},
  publisher={Taylor \& Francis}
}

@article{Fan01122001,
author = {Jianqing Fan and Runze Li},
title = {Variable Selection via Nonconcave Penalized Likelihood and its Oracle Properties},
journal = {Journal of the American Statistical Association},
volume = {96},
number = {456},
pages = {1348--1360},
year = {2001},
publisher = {ASA Website},
doi = {10.1198/016214501753382273},
URL = {    
        https://doi.org/10.1198/016214501753382273
},
eprint = {  
        https://doi.org/10.1198/016214501753382273
}

}

@article{thurman2015regularized,
  title={Regularized estimating equations for model selection of clustered spatial point processes},
  author={Thurman, Andrew L and Fu, Rao and Guan, Yongtao and Zhu, Jun},
  journal={Statistica Sinica},
  pages={173--188},
  year={2015},
  publisher={JSTOR}
}

@article{choiruddin2023adaptive,
  title={Adaptive lasso and Dantzig selector for spatial point processes intensity estimation},
  author={Choiruddin, Achmad and Coeurjolly, Jean-Fran{\c{c}}ois and Letu{\'e}, Fr{\'e}d{\'e}rique},
  journal={Bernoulli},
  volume={29},
  number={3},
  pages={1849--1876},
  year={2023},
  publisher={Bernoulli Society for Mathematical Statistics and Probability}
}

@article{mohler2019reducing,
  title={Reducing bias in estimates for the law of crime concentration},
  author={Mohler, George and Brantingham, P Jeffrey and Carter, Jeremy and Short, Martin B},
  journal={Journal of quantitative criminology},
  volume={35},
  number={4},
  pages={747--765},
  year={2019},
  publisher={Springer}
}

@article{heckman1991identifying,
  title={Identifying the hand of past: Distinguishing state dependence from heterogeneity},
  author={Heckman, James J},
  journal={The American Economic Review},
  volume={81},
  number={2},
  pages={75--79},
  year={1991},
  publisher={JSTOR}
}

@book{diggle2013statistical,
  title={Statistical analysis of spatial and spatio-temporal point patterns},
  author={Diggle, Peter J},
  year={2013},
  publisher={CRC press}
}

@article{mohler2011self,
  title={Self-exciting point process modeling of crime},
  author={Mohler, George O and Short, Martin B and Brantingham, P Jeffrey and Schoenberg, Frederic Paik and Tita, George E},
  journal={Journal of the american statistical association},
  volume={106},
  number={493},
  pages={100--108},
  year={2011},
  publisher={Taylor \& Francis}
}

@article{mohler2014marked,
  title={Marked point process hotspot maps for homicide and gun crime prediction in Chicago},
  author={Mohler, George},
  journal={International Journal of Forecasting},
  volume={30},
  number={3},
  pages={491--497},
  year={2014},
  publisher={Elsevier}
}

@article{reinhart2018self,
  title={Self-exciting point processes with spatial covariates: modelling the dynamics of crime},
  author={Reinhart, Alex and Greenhouse, Joel},
  journal={Journal of the Royal Statistical Society Series C: Applied Statistics},
  volume={67},
  number={5},
  pages={1305--1329},
  year={2018},
  publisher={Oxford University Press}
}

@incollection{andresen2012spatial,
  title={Spatial heterogeneity in crime analysis},
  author={Andresen, Martin A and Malleson, Nicolas},
  booktitle={Crime modeling and mapping using geospatial technologies},
  pages={3--23},
  year={2012},
  publisher={Springer}
}

@article{chen2025score,
  title={Score-based spatial-temporal point process for traffic accident prediction},
  author={Chen, Kehua and Luo, Yuhao and Zhu, Meixin and Wang, Xiaomeng and Wang, Hongcheng and Yang, Hai},
  journal={IEEE Transactions on Intelligent Transportation Systems},
  year={2025},
  publisher={IEEE}
}

@article{park2021investigating,
  title={Investigating clustering and violence interruption in gang-related violent crime data using spatial--temporal point processes with covariates},
  author={Park, Junhyung and Schoenberg, Frederic Paik and Bertozzi, Andrea L and Brantingham, P Jeffrey},
  journal={Journal of the American Statistical Association},
  volume={116},
  number={536},
  pages={1674--1687},
  year={2021},
  publisher={Taylor \& Francis}
}

@article{waagepetersen2007estimating,
  title={An estimating function approach to inference for inhomogeneous Neyman--Scott processes},
  author={Waagepetersen, Rasmus Plenge},
  journal={Biometrics},
  volume={63},
  number={1},
  pages={252--258},
  year={2007},
  publisher={Oxford University Press}
}

@article{waagepetersen2008estimating,
  title={Estimating functions for inhomogeneous spatial point processes with incomplete covariate data},
  author={Waagepetersen, Rasmus},
  journal={Biometrika},
  volume={95},
  number={2},
  pages={351--363},
  year={2008},
  publisher={Oxford University Press}
}

@article{waagepetersen2009two,
  title={Two-step estimation for inhomogeneous spatial point processes},
  author={Waagepetersen, Rasmus and Guan, Yongtao},
  journal={Journal of the Royal Statistical Society Series B: Statistical Methodology},
  volume={71},
  number={3},
  pages={685--702},
  year={2009},
  publisher={Oxford University Press}
}

@article{yue2015variable,
  title={Variable selection for inhomogeneous spatial point process models},
  author={Yue, Yu and Loh, Ji Meng},
  journal={Canadian Journal of Statistics},
  volume={43},
  number={2},
  pages={288--305},
  year={2015},
  publisher={Wiley Online Library}
}

@article{thurman2014variable,
  title={Variable selection for spatial Poisson point processes via a regularization method},
  author={Thurman, Andrew L and Zhu, Jun},
  journal={Statistical Methodology},
  volume={17},
  pages={113--125},
  year={2014},
  publisher={Elsevier}
}

@article{baddeley2014logistic,
  title={Logistic regression for spatial Gibbs point processes},
  author={Baddeley, Adrian and Coeurjolly, Jean-Fran{\c{c}}ois and Rubak, Ege and Waagepetersen, Rasmus},
  journal={Biometrika},
  volume={101},
  number={2},
  pages={377--392},
  year={2014},
  publisher={Oxford University Press}
}

@article{choiruddin2018convex,
author = {Achmad Choiruddin and Jean-Fran{\c{c}}ois Coeurjolly and Fr{\'e}d{\'e}rique Letu{\'e}},
title = {{Convex and non-convex regularization methods for spatial point processes intensity estimation}},
volume = {12},
journal = {Electronic Journal of Statistics},
number = {1},
publisher = {Institute of Mathematical Statistics and Bernoulli Society},
pages = {1210 -- 1255},
year = {2018},
doi = {10.1214/18-EJS1408}
}

@article{renner2013equivalence,
  title={Equivalence of MAXENT and Poisson point process models for species distribution modeling in ecology},
  author={Renner, Ian W and Warton, David I},
  journal={Biometrics},
  volume={69},
  number={1},
  pages={274--281},
  year={2013},
  publisher={Oxford University Press}
}

@article{baddeley2017local,
  title={Local composite likelihood for spatial point processes},
  author={Baddeley, Adrian},
  journal={Spatial Statistics},
  volume={22},
  pages={261--295},
  year={2017},
  publisher={Elsevier}
}

@article{coeurjolly2022convolution,
  title={A convolution type model for the intensity of spatial point processes applied to eye-movement data},
  author={Coeurjolly, Jean-Fran{\c{c}}ois and Cuevas-Pacheco, Francisco and Descary, Marie-H{\'e}l{\`e}ne},
  journal={Spatial Statistics},
  volume={51},
  pages={100651},
  year={2022},
  publisher={Elsevier}
}

@article{yin2021fused,
  title={Fused spatial point process intensity estimation with varying coefficients on complex constrained domains},
  author={Yin, Lihao and Sang, Huiyan},
  journal={Spatial Statistics},
  volume={46},
  pages={100547},
  year={2021},
  publisher={Elsevier}
}

@article{van2025non,
  title={A non-homogeneous alternating renewal process model for interval censoring},
  author={van Lieshout, MNM and Markwitz, RL},
  journal={Journal of Applied Probability},
  volume={62},
  number={2},
  pages={494--515},
  year={2025},
  publisher={Cambridge University Press}
}

@article{biscio2025nonparametric,
  title={Nonparametric intensity estimation of spatial point processes by random forests},
  author={Biscio, Christophe and Lavancier, Fr{\'e}d{\'e}ric},
  journal={arXiv preprint arXiv:2511.09307},
  year={2025}
}

@inproceedings{flaxman2017poisson,
  title={Poisson intensity estimation with reproducing kernels},
  author={Flaxman, Seth and Teh, Yee Whye and Sejdinovic, Dino},
  booktitle={Artificial Intelligence and Statistics},
  pages={270--279},
  year={2017},
  organization={PMLR}
}

@article{kim2022fast,
  title={Fast Bayesian estimation of point process intensity as function of covariates},
  author={Kim, Hideaki and Asami, Taichi and Toda, Hiroyuki},
  journal={Advances in Neural Information Processing Systems},
  volume={35},
  pages={25711--25724},
  year={2022}
}

@article{lu2026xgboostpp,
  title={XGBoostPP: Tree-based estimation of point process intensity functions},
  author={Lu, Changqing and Guan, Yongtao and van Lieshout, Marie-Colette and Xu, Ganggang},
  journal={Journal of Computational and Graphical Statistics},
  volume={35},
  number={1},
  pages={229--240},
  year={2026},
  publisher={Taylor \& Francis}
}

@book{mallat1999wavelet,
  title={A wavelet tour of signal processing},
  author={Mallat, St{\'e}phane},
  year={1999},
  publisher={Elsevier}
}

@article{choiruddin2021information,
  title={Information criteria for inhomogeneous spatial point processes},
  author={Choiruddin, Achmad and Coeurjolly, Jean-Fran{\c{c}}ois and Waagepetersen, Rasmus},
  journal={Australian \& New Zealand Journal of Statistics},
  volume={63},
  number={1},
  pages={119--143},
  year={2021},
  publisher={Wiley Online Library}
}

@incollection{brantingham2017environment,
  title={Environment, routine, and situation: Toward a pattern theory of crime},
  author={Brantingham, Patricia L and Brantingham, Paul J},
  booktitle={Routine activity and rational choice},
  pages={259--294},
  year={2017},
  publisher={Routledge}
}
\end{document}